\documentclass[preprint,showpacs,preprintnumbers,amsmath,amssymb,showkeys]{revtex4}

\usepackage{graphicx}
\usepackage{dcolumn}
\usepackage{bm}
\usepackage{braket}
\usepackage{enumitem}
\usepackage[utf8]{inputenc}
\usepackage{amsmath}
\usepackage{amssymb}
\usepackage{amsthm}

\begin{document}

\noindent

\preprint{}

\title{Quantum entanglement as an extremal Kirkwood-Dirac nonreality}

\author{Agung Budiyono}
\email{agungbymlati@gmail.com}
\affiliation{Research Center for Quantum Physics, National Research and Innovation Agency, South Tangerang 15314, Republic of Indonesia} 
\affiliation{Research Center for Nanoscience and Nanotechnology, Bandung Institute of Technology, Bandung, 40132, Republic of Indonesia}
\affiliation{Department of Engineering Physics, Bandung Institute of Technology, Bandung, 40132, Republic of Indonesia} 

\date{\today}

\begin{abstract} 

Understanding the relationship between various different forms of nonclassicality and their resource character is of great importance in quantum foundation and quantum information. Here, we discuss a quantitative link between quantum entanglement and the anomalous or nonclassical nonreal values of Kirkwood-Dirac (KD) quasiprobability, in a bipartite setting. We first construct an entanglement monotone for a pure bipartite state based on the nonreality of the KD quasiprobability defined over a pair of orthonormal bases in which one of them is a product, and optimizations over these bases. It admits a closed expression as a Schur-concave function of the state of the subsystem having a form of nonadditive quantum entropy. We then construct a bipartite entanglement monotone for generic quantum states using the convex roof extension. Its normalized value is upper bounded by the concurrence of formation, and for two-qubit systems, they are equal. We also derive lower and upper bounds in terms of different forms of uncertainty in the subsystem quantified, respectively, by an extremal trace-norm asymmetry and a nonadditive quantum entropy. The entanglement monotone can be expressed as the minimum total state disturbance due to a nonselective local binary measurement. Finally, we discuss its estimation using weak value measurement and classical optimization, and its connection with strange weak value and quantum contextuality. 

\end{abstract} 

\maketitle       


\section{Introduction} 

Quantum entanglement \cite{EPR paper}, Schr\"odinger famously said, is the characteristic trait of quantum mechanics which enforces its entire departure from the classical line of thought \cite{Schroedinger entanglement paper 1,Schroedinger entanglement paper 2}. It defies any classical explanation through e.g. the violation of the Bell inequality \cite{Bell theorem,CHSH inequality,Aspect experiment,Hensen loophole free Bell test,Giustina loophole free Bell test,Shalm loophole free Bell test}. The advent of the quantum information theory and quantum technology \cite{Nielsen-Chuang book} in the last decades has led to the recognition of entanglement as a key nonclassical resource as real as energy \cite{Horodecki entanglement review,Plenio review}. That is, entanglement enables various schemes of information processing which are significantly more efficient and secure than their classical counterparts or cannot be achieved using classical physics such as: quantum cryptography \cite{Ekert quantum cryptography using entanglement} and quantum random number generator \cite{Bierhorst experiment QRNG}, quantum dense coding \cite{Bennett quantum dense coding}, quantum teleportation \cite{Bennett quantum teleportation}, quantum metrology \cite{Giovannetti quantum metrology review} and quantum computation \cite{Nielsen-Chuang book}. Recent ramification of the concept of entanglement also showed that it links traditionally disparate fields of physics: AMO, condensed matter, quantum gravity, and high-energy physics \cite{Lewis-Swan entanglement underlying various physical phenomena in traditionally separate fields}. The above facts suggest that different approaches for the characterization and quantification of entanglement from various perspectives, in particular those that can be implemented using the near-term quantum hardware, are desirable to reveal its rich multi-faceted features, to understand its deep meaning and its relationship with other forms of nonclassicality, and to search for a new avenue of applications.  

On the other hand, since the early days, there have been significant efforts to better understand the statistics associated with a pair of noncommuting quantum measurements as some sort of modification of classical probability. A powerful method in this line of approach is by using the mathematical concept of quasiprobability which can be seen as a quantum analog of phase space probability distribution of classical statistical mechanics \cite{Wigner function,Lee quasiprobability review,Ferrie quasiprobability representation}. Quantum noncommutativity implies that no quasiprobability that is convex-linear in the density operator, can satisfy all the Kolmogorov axioms of classical probability theory, i.e., real-nonnegative with correct marginal probabilities \cite{Lostaglio KD quasiprobability and quantum fluctuation}. Hence, it is natural to ask if such a deviation from classical probability can be used to access the quantum entanglement which also arises partly due to noncommutativity. To this end, let us mention that, building on the observation in Refs. \cite{Sanpera entanglement quasiprobability,Vidal entanglement quasiprobability}, a specific quasiprobability called entanglement quasiprobability was designed so that its negative values witness entanglement \cite{Sperling entanglement by negative quasiprobability,Sperling entanglement quasiprobability of squeezed light,Sperling entanglement from negativity of certain quasiprobability: finite system}. 

In this article we work with Kirkwood-Dirac (KD) quasiprobability \cite{Kirkwood quasiprobability,Dirac quasiprobability,Barut KD quasiprobability,Schmid KD representation for all quantum operations}, a specific quasiprobability that is convex-linear in the density operator. Kirkwood-Dirac quasiprobability yields correct marginal probabilities, but it may assume nonreal values and its real part maybe negative or larger than one, manifesting the noncommutativity among the quantum state and the two defining bases. Recently, such anomalous or nonclassical values of the KD quasiprobability have been shown to play crucial roles in different areas of quantum science and technology \cite{Lostaglio KD quasiprobability and quantum fluctuation,Allahverdyan TBMH as quasiprobability distribution of work,Lostaglio TBMH quasiprobability fluctuation theorem contextuality,Levy quasiprobability distribution for heat fluctuation in quantum regime,Alonso KD quasiprobability witnesses quantum scrambling,Halpern quasiprobability and information scrambling,Arvidsson-Shukur quantum advantage in postselected metrology,Lupu-Gladstein negativity enhanced quantum phase estimation 2022,Pusey negative TBMH quasiprobability and contextuality,Kunjwal contextuality of non-real weak value,Lostaglio contextuality in quantum linear response,Agung KD-nonreality coherence,Agung KD-nonclassicality coherence,Agung KD-nonreality general quantum correlation,Agung trade-off relations for KD quantumness,Agung separation of measurement uncertainty into quantum and classical parts}. Here, we show that the nonreality of the KD quasiprobability can indeed be used to characterize and quantify bipartite entanglement in a quantum state on a finite-dimensional Hilbert space by devising an entanglement monotone. The intuition behind the approach is that the nonreality of the KD quasiprobability of a bipartite state over a pair of orthonormal bases, in which one of them is a product, and suitably optimized over these bases, captures the noncommutativity between the bipartite state and any product basis describing local measurement, which in turn gives access to the entanglement of the bipartite state. Hence, our entanglement monotone captures the genuine quantum correlation arising from the incompatibility between the state and any local measurement.  

We first show that for any pure bipartite state, such extremal nonreality of the KD quasiprobability has a closed expression as a Schur-concave function of the quantum state of the subsystem having the form of a nonadditive quantum entropy, and hence it is an entanglement monotone \cite{Vidal entanglement monotone}. We then construct a measure of entanglement for a generic state using the convex roof extension. Its normalized value gives a lower bound to the concurrence of formation and they are equal for any two-qubit state. Moreover, it is also upper bounded by the purity of the state of the system quantified by certain nonadditive quantum entropy and lower bounded by certain extremal trace-norm asymmetry relative to the group of translation unitaries generated by any local Hermitian operator. In this sense, it is lower and upper bounded by different forms of uncertainty in the subsystem. We derive a geometrical interpretation in terms of the minimum trace distance between the original state and the final state after a specific nonselective local binary measurement. We then sketch a variational scheme for the estimation of the entanglement measure for an unknown quantum state using weak value measurement \cite{Aharonov weak value,Aharonov-Daniel book,Wiseman weak value,Lundeen complex weak value,Jozsa complex weak value,Lundeen measurement of KD distribution,Salvail direct measurement KD distribution,Bamber measurement of KD distribution,Thekkadath measurement of density matrix,Johansen quantum state from successive projective measurement,Hernandez-Gomez experimental observation of TBMH negativity,Wagner measuring weak values and KD quasiprobability,Vallone strong measurement to reconstruct quantum wave function,Cohen estimating of weak value with strong measurements,Chiribella dimension independent weak value estimation independent} and classical optimization, and discuss its relation to the strange weak value and proof of quantum contextuality \cite{Spekkens generalized quantum contextuality}.

\section{Preliminary}

\subsection{Entanglement monotone and noncommutativity} 

Consider a bipartite system AB with a finite-dimensional Hilbert space $\mathcal{H}_{AB}=\mathcal{H}_A\otimes\mathcal{H}_B$, where $\mathcal{H}_{A(B)}$ is the Hilbert space of the subsystem A(B). A bipartite state represented by a density operator $\varrho_{AB}$ on $\mathcal{H}_{AB}$ is separable if and only if it can be prepared by a set of local operations and classical communication (LOCC). Any separable bipartite state can thus be expressed as $\varrho_{AB}=\sum_kp_k\varrho_A^k\otimes\varrho_B^k$, where $\{p_k\}$ is a set of normalized probabilities and $\{\varrho_{A(B)}^k\}$ is a set of density operators on $\mathcal{H}_{A(B)}$ \cite{Werner separable state}. Otherwise, the state is entangled. Hence, entanglement is a genuine quantum correlation which cannot be created using LOCC alone. It is therefore reasonable to impose on any functional of the bipartite state returning a nonnegative real number, i.e., $\varrho_{AB}\mapsto E(\varrho_{AB})$, to be regarded as a monotone (measure) of entanglement in the state $\varrho_{AB}$, the following requirements: 1) monotonicity, i.e., nonincreasing on average under LOCC, and 2) convexity, i.e., nonincreasing under mixing of states \cite{Horodecki entanglement review}. 

Different approaches have been proposed to construct entanglement monotones satisfying the above plausible requirements. A powerful and elegant method to construct a bipartite entanglement monotone was suggested by Vidal \cite{Vidal entanglement monotone}. First, one starts by developing an entanglement monotone for a pure state $\varrho_{AB}=\ket{\psi}\bra{\psi}_{AB}$. Let $\varrho_{A(B)}={\rm Tr}_{B(A)}\{\ket{\psi}\bra{\psi}_{AB}\}$ be the reduced density operator, i.e., the quantum state of the subsystem A(B). Then, any mapping of $\varrho_{AB}=\ket{\psi}\bra{\psi}_{AB}$ to a nonnegative real number which (i) depends only on $\varrho_A$ ($\varrho_B$), (ii) invariant under local unitary transformation: $\varrho_{A(B)}\mapsto U_{A(B)}\varrho_{A(B)}U_{A(B)}^{\dagger}$, with $U_{A(B)}$ any unitary locally operating on subsystem A(B), and (iii) concave with respect to $\varrho_{A(B)}$, can be used as a bipartite entanglement monotone for pure states. The above set of conditions is equivalent to the requirement that it is a function only of the Schmidt coefficients of the pure bipartite state, and is monotonic under majorization of the associated Schmidt probability vector. Namely, it must be a Schur-concave function of the Schmidt probability vector. This captures the intuition that for a pure state, entanglement manifests in the local uncertainty of each subsystem which is symmetric under the exchange of the subsystem, and cannot increase under any LOCC. The entanglement monotone for pure states can then be generalized to the whole set of states via the convex roof construction. 

There are infinitely many quantities which satisfy the above conditions for an entanglement monotone. In the present work, we pick a specific form of such an entanglement monotone which can be obtained via the nonreality of a certain KD quasiprobability suitably optimized. Our main goal is thus linking the notion of entanglement and anomalous values of the KD quasiprobability, two seemingly separate nonclassical features of quantum mechanics. Such an alternative approach is important to unveil yet unexplored facets of entanglement which may stimulate fresh applications in either quantum foundation or quantum information theory.  

We first recall that, according to the Schmidt decomposition, an arbitrary bipartite pure state can always be expressed as $\ket{\psi}_{AB}=\sum_{j=1}^d\sqrt{\lambda_j}\ket{\mu^j}_A\ket{\nu^j}_B$, where $\lambda_j\in\mathbb{R}^+$, $\sum_{j=1}^d\lambda_j=1$ with $d$ the Schmidt rank, and $\{\ket{\mu^j}_A\}$ and $\{\ket{\nu^k}_B\}$ are two sets of orthonormal vectors in $\mathcal{H}_A$ and $\mathcal{H}_B$, respectively. A pure bipartite state is thus nonfactorizable or entangled if and only if the Schmidt rank is larger than 1, i.e., $d\ge 2$. Observe then that for a bipartite entangled pure state, any nonselective local projective measurement, e.g., on the side of A described by a local (product) projection-valued measure (PVM) $\{\ket{a}\bra{a}_A\otimes\mathbb{I}_B\}$ on $\mathcal{H}_{AB}$, unavoidably disturbs the global state, i.e., $\sum_a(\ket{a}\bra{a}_A\otimes\mathbb{I}_B)\ket{\psi}\bra{\psi}_{AB}(\ket{a}\bra{a}_A\otimes\mathbb{I}_B)\neq\ket{\psi}\bra{\psi}_{AB}$, for all choices of rank-1 PVM $\{\ket{a}\bra{a}_A\}$ on $\mathcal{H}_A$. Hence, in this case, any local measurement described by PVM $\{\ket{a}\bra{a}_A\otimes\mathbb{I}_B\}$ does not commute with the bipartite state, i.e., $[\ket{a}\bra{a}_A\otimes\mathbb{I}_B,\ket{\psi}\bra{\psi}_{AB}]:=(\ket{a}\bra{a}_A\otimes\mathbb{I}_B)\ket{\psi}\bra{\psi}_{AB}-\ket{\psi}\bra{\psi}_{AB}(\ket{a}\bra{a}_A\otimes\mathbb{I}_B)\neq 0$, for some $a$. See also the proof of Property 1 below in Appendix \ref{Proof of properties ENR1-ENR3}. We will later exploit this noncommutativity to characterize and quantify entanglement in a generic bipartite state, and indeed the entanglement measure thus obtained can be characterized geometrically in terms of the distance between the original state and state after any local measurement. 

\subsection{Kirkwood-Dirac quasiprobability for bipartite systems} 

Consider a system with a finite-dimensional Hilbert space $\mathcal{H}$. Choose a pair of orthonormal bases $\{\ket{x}\}$ and $\{\ket{y}\}$ of $\mathcal{H}$. The KD quasiprobability associated with a quantum state $\varrho$ on $\mathcal{H}$ relative to the pair of orthonormal bases is defined as ${\rm Pr}_{\rm KD}(x,y|\varrho):={\rm Tr}\{\Pi_y\Pi_x\varrho\}$ \cite{Kirkwood quasiprobability,Dirac quasiprobability,Barut KD quasiprobability}, where $\Pi_i:=\ket{i}\bra{i}$, $i=x,y$, is a projector on the subset of the Hilbert space spanned by $\ket{i}$. In the following, we refer to $\{\ket{x}\}$ and $\{\ket{y}\}$ as the first and the second defining bases, respectively. Note that $\{\Pi_i\}$, $i=x,y$, comprises a rank-$1$ orthogonal PVM describing a projective measurement with the probability to get outcome $i$ given by the Born's rule as ${\rm Pr}(i|\varrho)={\rm Tr}\{\Pi_i\varrho\}$. The KD quasiprobability returns correct marginal probabilities, i.e., $\sum_i{\rm Pr}_{\rm KD}(x,y|\varrho)={\rm Pr}(j|\varrho)$, $j\neq i$, $i,j=x,y$, and thus normalized: $\sum_{x,y}{\rm Pr}_{\rm KD}(x,y|\varrho)=1$, as for the conventional probability. However, unlike the latter, KD quasiprobability may assume nonreal values, and its real part may be negative or larger than 1. Such nonclassical values manifest the noncommutativity among the state $\varrho$ and the pair of the defining PVMs $\{\Pi_x\}$ and $\{\Pi_y\}$, i.e., assuming two of them commute renders the KD quasiprobability real and nonnegative. On the other hand, given a KD quasiprobability ${\rm Pr}_{\rm KD}(x,y|\varrho)$ over a pair of defining orthonormal bases $\{\ket{x}\}$ and $\{\ket{y}\}$ with $\braket{y|x}\neq 0$ for all $x,y$, one can recover the associated quantum state as $\varrho=\sum_{x,y}{\rm Pr}_{\rm KD}(x,y|\varrho)\frac{\ket{x}\bra{y}}{\braket{y|x}}$. In this sense, KD quasiprobability can provide an informationally complete representation of a quantum state. The above observation prompts the question of how the nonclassical values of the KD quasiprobability encode the nonclassical aspects of quantum mechanics stemming from the noncommutativity between the quantum state and measurement operators \cite{Agung KD-nonreality coherence,Agung KD-nonclassicality coherence,Agung KD-nonreality general quantum correlation,Agung trade-off relations for KD quantumness,Pusey negative TBMH quasiprobability and contextuality,Kunjwal contextuality of non-real weak value,Lostaglio contextuality in quantum linear response}. 

Now, consider again a bipartite quantum system AB with a finite-dimensional Hilbert space $\mathcal{H}_{AB}$. For a reason that will be clear as we proceed, we choose as the first basis for defining the KD quasiprobability, an orthonormal product basis of $\mathcal{H}_{AB}$ as: $\{\ket{x_A,x_B}_{AB}:=\ket{x_A}_A\otimes\ket{x_B}_B\}$, where $\{\ket{x_A}_A\}$ and $\{\ket{x_B}_B\}$ are orthonormal bases of $\mathcal{H}_A$ and $\mathcal{H}_B$, respectively. On the other hand, as for the second basis for defining the KD quasiprobability, we choose any orthonormal basis $\{\ket{y_{AB}}_{AB}\}$ of the Hilbert space $\mathcal{H}_{AB}$. Then, the KD quasiprobability associated with a bipartite state $\varrho_{AB}$ on $\mathcal{H}_{AB}$ relative to the pair of the orthonormal bases $\{\ket{x_A,x_B}_{AB}\}$ and $\{\ket{y_{AB}}_{AB}\}$ reads  ${\rm Pr}_{\rm KD}(x_A,x_B;y_{AB}|\varrho_{AB}):={\rm Tr}\{\Pi_{y_{AB}}(\Pi_{x_A}\otimes\Pi_{x_B})\varrho_{AB}\}$. Moreover, summing over one of the variables associated with the first orthonormal product basis, e.g., over $x_B$, we obtain
\begin{eqnarray}
{\rm Pr}_{\rm KD}(x_A;y_{AB}|\varrho_{AB})&:=&\sum_{x_B}{\rm Pr}_{\rm KD}(x_A,x_B;y_{AB}|\varrho_{AB})\nonumber\\
&=&{\rm Tr}\{\Pi_{y_{AB}}(\Pi_{x_A}\otimes\mathbb{I}_B)\varrho_{AB}\}. 
\label{KD quasiprobability with one variable for the first orthonormal product basis}
\end{eqnarray}
One can check that ${\rm Pr}_{\rm KD}(x_A;y_{AB}|\varrho_{AB})$ defined in Eq. (\ref{KD quasiprobability with one variable for the first orthonormal product basis}) too has correct marginal probabilities, and thus is normalized. However, it may take nonreal values, and its real part may be negative or larger than 1. We thus also refer to it as KD quasiprobability. We stress that the nonclassical values of ${\rm Pr}_{\rm KD}(x_A;y_{AB}|\varrho_{AB})$ manifest the noncommutativity among the bipartite state $\varrho_{AB}$, the first defining local PVM $\{\Pi_{x_A}\otimes\mathbb{I}_B\}$, and the second defining PVM $\{\Pi_{y_{AB}}\}$. This fact suggests that the nonclassical values of the KD quasiprobability ${\rm Pr}_{\rm KD}(x_A;y_{AB}|\varrho_{AB})$ is closely linked to the bipartite entanglement in $\varrho_{AB}$ recalling our previous observation that, for pure states, bipartite entanglement also captures the noncommutativity between the state $\varrho_{AB}=\ket{\psi}\bra{\psi}_{AB}$ and any local measurement described by the PVM $\{\Pi_{x_A}\otimes\mathbb{I}_B\}$.  

\section{Monotone for pure state bipartite entanglement from Kirkwood-Dirac nonreality} 

We wish to use the nonreality of the KD quasiprobability ${\rm Pr}_{\rm KD}(x_A;y_{AB}|\ket{\psi}\bra{\psi}_{AB})$ defined in Eq. (\ref{KD quasiprobability with one variable for the first orthonormal product basis}), suitably optimized so that it captures the noncommutativity between the state $\ket{\psi}\bra{\psi}_{AB}$ and any local measurement described by the PVM $\{\Pi_{x_A}\otimes \mathbb{I}_B\}$, to characterize and quantify the bipartite entanglement in the pure state $\ket{\psi}_{AB}$. A natural way to quantify the nonreality of ${\rm Pr}_{\rm KD}(x_A;y_{AB}|\ket{\psi}\bra{\psi}_{AB})$ is by using the $l_1$-norm of the total nonreal values in ${\rm Pr}_{\rm KD}(x_A;y_{AB}|\ket{\psi}\bra{\psi}_{AB})$. We thus introduce the following quantity.\\
{\bf Definition 1}. Given a pure bipartite state $\ket{\psi}_{AB}$ in a finite-dimensional Hilbert space $\mathcal{H}_{AB}$, define a quantity that maps the state to a nonnegative real number, as follows:
\begin{eqnarray}
&&E_{\rm KD}^{\rm NRe}(\ket{\psi}\bra{\psi}_{AB})\nonumber\\
&=&\inf_{\{\ket{x_A}\}\in\mathcal{B}_{\rm o}(\mathcal{H}_A)}\sum_{x_A}\sup_{\{\ket{y_{AB}}\}\in\mathcal{B}_{\rm o}(\mathcal{H}_{AB})}\sum_{y_{AB}}|{\rm Im}({\rm Pr}_{\rm KD}(x_A;y_{AB}|\ket{\psi}\bra{\psi}_{AB})|\nonumber\\
&=&\inf_{\{\Pi_{x_A}\}\in\mathcal{M}_{\rm r1PVM}(\mathcal{H}_A)}\sum_{x_A}\sup_{\{\ket{y_{AB}}\}\in\mathcal{B}_{\rm o}(\mathcal{H}_{AB})}\sum_{y_{AB}}\frac{1}{2}\big|\braket{y_{AB}|[(\Pi_{x_A}\otimes \mathbb{I}_B),\ket{\psi}\bra{\psi}_{AB}]|y_{AB}}\big|.  
\label{KD-nonreality entanglement for pure bipartite state} 
\end{eqnarray}
Here, the supremum is taken over the set $\mathcal{B}_{\rm o}(\mathcal{H}_{AB})$ of all orthonormal bases of $\mathcal{H}_{AB}$, the infimum is taken over the set $\mathcal{B}_{\rm o}(\mathcal{H}_A)$ of orthonormal bases of $\mathcal{H}_A$ of subsystem A, and $\mathcal{M}_{\rm r1PVM}(\mathcal{H}_A)$ is the set of all rank-1 PVMs on $\mathcal{H}_A$ associated with the set $\mathcal{B}_{\rm o}(\mathcal{H}_A)$ of orthonormal bases of $\mathcal{H}_A$. Moreover, the evaluation of the infimum is done after the search for the supremum. 

A similar quantity for a generic state $\varrho_{AB}$, where the supremum over the second basis is taken over all orthonormal product bases of $\mathcal{H}_{AB}$, was introduced in an earlier work as a quantifier of discord-like general quantum correlation \cite{Agung KD-nonreality general quantum correlation}. Below we show that $E_{\rm KD}^{\rm NRe}(\ket{\psi}\bra{\psi}_{AB})$ defined in Eq. (\ref{KD-nonreality entanglement for pure bipartite state}) in which the second basis is not restricted to a product basis, is indeed a bipartite entanglement monotone for the pure state $\ket{\psi}_{AB}$. 

We first prove in Appendix \ref{Proof of properties ENR1-ENR3} that $E_{\rm KD}^{\rm NRe}(\ket{\psi}\bra{\psi}_{AB})$ possesses the following properties:  \\ 
{\bf Property 1}. $E_{\rm KD}^{\rm NRe}(\ket{\psi}\bra{\psi}_{AB})$ is vanishing if and only if the pure state $\ket{\psi}_{AB}$ is unentangled or factorizable. \\
{\bf Property 2}. $E_{\rm KD}^{\rm NRe}(\ket{\psi}\bra{\psi}_{AB})$ is invariant under arbitrary local unitary transformation, i.e., $E_{\rm KD}^{\rm NRe}\big((U_A\otimes U_B)\ket{\psi}\bra{\psi}_{AB}(U_A\otimes U_B)^{\dagger}\big)=E_{\rm KD}^{\rm NRe}(\ket{\psi}\bra{\psi}_{AB})$, where $U_{A(B)}$ is any unitary operator applying locally on subsystem A(B).  \\

Next, we show that $E_{\rm KD}^{\rm NRe}(\ket{\psi}\bra{\psi}_{AB})$ defined in Eq. (\ref{KD-nonreality entanglement for pure bipartite state}) depends only on the reduced density operator $\varrho_A={\rm Tr}_B\{\ket{\psi}\bra{\psi}_{AB}\}$, i.e., the state of subsystem A, or symmetrically, only on the state $\varrho_B={\rm Tr}_A\{\ket{\psi}\bra{\psi}_{AB}\}$ of subsystem B. First, we have the following result. \\
{\bf Proposition 1}. $E_{\rm KD}^{\rm NRe}(\ket{\psi}\bra{\psi}_{AB})$ is equal to the minimum of the total standard deviation of the PVM $\{\Pi_{x_A}\otimes\mathbb{I}_B\}$ in the state $\ket{\psi}_{AB}$ as
\begin{eqnarray}
&&E_{\rm KD}^{\rm NRe}(\ket{\psi}\bra{\psi}_{AB})\nonumber\\
\label{KD entanglement versus local quantum uncertainty step 1}
&=&\inf_{\{\Pi_{x_A}\}\in\mathcal{M}_{\rm r1PVM}(\mathcal{H}_A)}\sum_{x_A}\|[\ket{\psi}\bra{\psi}_{AB},(\Pi_{x_A}\otimes\mathbb{I}_B)]\|_1/2\\
\label{KD entanglement versus local quantum uncertainty step 2}
&=&\inf_{\{\Pi_{x_A}\}\in\mathcal{M}_{\rm r1PVM}(\mathcal{H}_A)}\sum_{x_A}\Delta_{\Pi_{x_A}}(\varrho_A)\\
\label{KD entanglement versus local quantum uncertainty step 3}
&=&\inf_{\{\Pi_{x_A}\}\in\mathcal{M}_{\rm r1PVM}(\mathcal{H}_A)}\sum_{x_A}\big({\rm Pr}(x_A|\varrho_A)-{\rm Pr}(x_A|\varrho_A)^2\big)^{1/2}\\
\label{KD entanglement versus local quantum uncertainty step 4}
&\le&\sqrt{d-1}.
\label{equality between pure state KD entanglement and total local uncertainty}
\end{eqnarray}
Here, $\|O\|_1={\rm Tr}\{|O|\}$, with $|O|=\sqrt{OO^{\dagger}}$, is the trace norm of an operator $O$, and $\Delta_O(\varrho)^2={\rm Tr}\{O^2\varrho\}-{\rm Tr}\{O\varrho\}^2$ is the quantum variance of a Hermitian operator $O$ in the state $\varrho$. Moreover, the inequality in Eq. (\ref{KD entanglement versus local quantum uncertainty step 4}) becomes an equality for a maximally entangled state, where $d$ is the Schmidt rank of the bipartite pure state. \\
{\bf Proof}. See Appendix \ref{Proof of first part of Proposition 1}. 

Equation (\ref{KD entanglement versus local quantum uncertainty step 2}) shows that $E_{\rm KD}^{\rm NRe}(\ket{\psi}\bra{\psi}_{AB})$ defined in Eq. (\ref{KD-nonreality entanglement for pure bipartite state}) quantifies a minimum total local uncertainty arising in all local projective measurement on subsystem A described by the rank-1 PVM $\{\Pi_{x_A}\}$ over the state $\varrho_A$ of subsystem A wherein the subsystem B is left untouched. It captures the intuition that when $E_{\rm KD}^{\rm NRe}(\ket{\psi}\bra{\psi}_{AB})$ is nonvanishing, even though the pure state $\ket{\psi}_{AB}$ of the bipartite system can be in principle perfectly determined, each subsystem nevertheless behaves randomly. One can further see that when the pure state is factorizable $\ket{\psi}_{AB}=\ket{\psi_A}_A\ket{\psi_B}_B$, one can always find a rank-1 PVM $\{\Pi_{x_A}\}$ whose elements include $\ket{\psi_A}\bra{\psi_A}$ so that $E_{\rm KD}^{\rm NRe}(\ket{\psi}\bra{\psi}_{AB})=\inf_{\{\Pi_{x_A}\}\in\mathcal{M}_{\rm r1PVM}(\mathcal{H}_A)}\sum_{x_A}\Delta_{\Pi_{x_A}}(\ket{\psi_A}\bra{\psi_{A}})=0$, as expected.  

One also observes in Eqs. (\ref{KD entanglement versus local quantum uncertainty step 2}) or (\ref{KD entanglement versus local quantum uncertainty step 3}) that $E_{\rm KD}^{\rm NRe}(\ket{\psi}\bra{\psi}_{AB})$ only depends on the state $\varrho_A$ of subsystem A. Since $E_{\rm KD}^{\rm NRe}(\ket{\psi}\bra{\psi}_{AB})$ is also invariant under any local unitary transformation as per Property 2, then it must depend only on the eigenvalues of $\varrho_A$. Hence, writing the bipartite pure state according to the Schmidt decomposition as $\ket{\psi}_{AB}=\sum_{j=1}^d\sqrt{\lambda_j}\ket{\mu^j}_A\ket{\nu^j}_B$, $\lambda_j\in\mathbb{R}^+$, $\sum_{j=1}^d\lambda_j=1$, $E_{\rm KD}^{\rm NRe}(\ket{\psi}\bra{\psi}_{AB})$ must only depend on the set of Schmidt coefficients: $\{\lambda_j^{1/2}\}$, $j=1,\dots,d$. This further means that it only depends symmetrically on $\varrho_B$. That is, it is invariant under the exchange of subsystems A and B, as claimed. Finally, we note that the right-hand side of Eq. (\ref{KD entanglement versus local quantum uncertainty step 3}), i.e., $\sum_{x_A}\big({\rm Pr}(x_A|\varrho_A)-{\rm Pr}(x_A|\varrho_A)^2)^{1/2}$ is invariant under the permutation of the elements of the set of the probabilities $\{{\rm Pr}(x_A|\varrho_A)\}$. Moreover, it is concave with respect to $\varrho_A$, so that the infimum over $\{\Pi_{x_A}\}\in\mathcal{M}_{\rm r1PVM}(\mathcal{H}_A)$ is a concave function of $\varrho_A$. Hence, $E_{\rm KD}^{\rm NRe}(\ket{\psi}\bra{\psi}_{AB})$ defined in Eq. (\ref{KD-nonreality entanglement for pure bipartite state}) fulfils  the sufficient requirements for a pure state bipartite entanglement monotone \cite{Vidal entanglement monotone}. We have thus the following theorem.\\
{\bf Theorem 1}. $E_{\rm KD}^{\rm NRe}(\ket{\psi}\bra{\psi}_{AB})$ defined in Eq. (\ref{KD-nonreality entanglement for pure bipartite state}) is a monotonic measure of bipartite entanglement for the pure state $\ket{\psi}_{AB}$.  

Below we refer to $E_{\rm KD}^{\rm NRe}(\ket{\psi}\bra{\psi}_{AB})$ as the KD-nonreality bipartite entanglement for the pure state $\ket{\psi}_{AB}$. From the fact that $E_{\rm KD}^{\rm NRe}(\ket{\psi}\bra{\psi}_{AB})$ is symmetric under the exchange of the subsystems A and B, it can also be expressed as $E_{\rm KD}^{\rm NRe}(\ket{\psi}\bra{\psi}_{AB})=\inf_{\{\ket{x_B}\}\in\mathcal{B}_{\rm o}(\mathcal{H}_B)}\sum_{x_B}\sup_{\{\ket{y_{AB}}\}\in\mathcal{B}_{\rm o}(\mathcal{H}_{AB})}\sum_{y_{AB}}\frac{1}{2}\big|\braket{y_{AB}|[(\mathbb{I}_A\otimes \Pi_{x_B}),\ket{\psi}\bra{\psi}_{AB}]|y_{AB}}\big|=\inf_{\{\Pi_{x_B}\}\in\mathcal{M}_{\rm r1PVM}(\mathcal{H}_B)}\sum_{x_B}\big({\rm Pr}(x_B|\varrho_B)-{\rm Pr}(x_B|\varrho_B)^2\big)^{1/2}$. The invariance of $E_{\rm KD}^{\rm NRe}(\ket{\psi}\bra{\psi}_{AB})$ under the exchange of A and B is physically expected since the bipartite entanglement is a form of quantum correlation between the subsystems A and B. 

Before proceeding, let us comment on the relation between the KD-nonreality bipartite entanglement $E_{\rm KD}^{\rm NRe}(\ket{\psi}\bra{\psi}_{AB})$ for the pure state $\ket{\psi}_{AB}$ defined in Eq. (\ref{KD-nonreality entanglement for pure bipartite state}) and a coherence quantifier, called KD-nonreality coherence, proposed previously in Ref. \cite{Agung KD-nonreality coherence}, which is defined also based on the nonreal values of KD quasiprobability. First, in Ref. \cite{Agung KD-nonreality coherence}, we defined KD-nonreality coherence in the state $\varrho_A$ relative to an orthonormal basis $\{\ket{x_A}\}$ of $\mathcal{H}_A$  as $C_{\rm KD}^{\rm NRe}(\varrho_A;\{\ket{x_A}\}):=\sum_{x_A}\sup_{\{\ket{y_A}\}\in\mathcal{B}_{\rm o}(\mathcal{H}_A)}\sum_{y_A}|{\rm Im}({\rm Pr}_{\rm KD}(x_A,y_A|\varrho_A)|$, where ${\rm Pr}_{\rm KD}(x_A,y_A|\varrho_A)={\rm Tr}\{\Pi_{y_A}\Pi_{x_A}\varrho_A\}$ is just the KD quasiprobability associated with the state $\varrho_A$ relative to the pair of orthonormal bases $\{\ket{x_A}\}$ and $\{\ket{y_A}\}$ of $\mathcal{H}_A$. As shown in Ref. \cite{Agung separation of measurement uncertainty into quantum and classical parts}, $C_{\rm KD}^{\rm NRe}(\varrho_A;\{\ket{x_A}\})$ is upper bounded by the term inside the infimum on the right-hand side of Eq. (\ref{KD entanglement versus local quantum uncertainty step 3}), i.e., 
\begin{eqnarray}
C_{\rm KD}^{\rm NRe}(\varrho_A;\{\ket{x_A}\})\le\sum_{x_A}\big({\rm Pr}(x_A|\varrho_A)-{\rm Pr}(x_A|\varrho_A)^2\big)^{1/2}. 
\label{KD nonreality coherence is upper bounded by S entropy}
\end{eqnarray}
Equation (\ref{KD entanglement versus local quantum uncertainty step 3}) thus shows that the KD-nonreality bipartite entanglement  $E_{\rm KD}^{\rm NRe}(\ket{\psi}\bra{\psi}_{AB})$ for the pure state $\ket{\psi}\bra{\psi}_{AB}$ is obtained as the minimum of the upper bound of the KD-nonreality coherence in the state $\varrho_A={\rm Tr}_B\{\ket{\psi}\bra{\psi}_{AB}\}$ of subsystem A over all reference orthonormal bases of $\mathcal{H}_A$. Note that when $\varrho_A$ is maximally mixed which corresponds to the a maximally entangled pure state $\ket{\psi}_{AB}$, the coherence on the left-hand side of Eq. (\ref{KD nonreality coherence is upper bounded by S entropy}) is vanishing while the local uncertainty in  subsystem A quantified by the right-hand side of Eq. (\ref{KD nonreality coherence is upper bounded by S entropy}) takes its maximum value. Hence, the pure state bipartite entanglement does not appear in the local coherence. Rather, it appears as the minimum of the local uncertainty in the subsystem quantified by the right-hand side of Eq. (\ref{KD nonreality coherence is upper bounded by S entropy}). 

Next, we derive a closed expression for $E_{\rm KD}^{\rm NRe}(\ket{\psi}\bra{\psi}_{AB})$ linking the KD-nonreality entanglement for pure bipartite state to a Schur-concave function of the Schmidt probability vector. \\
{\bf Proposition 2}. $E_{\rm KD}^{\rm NRe}(\ket{\psi}\bra{\psi}_{AB})$ defined in Eq. (\ref{KD-nonreality entanglement for pure bipartite state}) can be expressed in terms of the state of the subsystems $\varrho_{A(B)}={\rm Tr}_{B(A)}\{\ket{\psi}\bra{\psi}_{AB}\}$ as 
\begin{eqnarray}
E_{\rm KD}^{\rm NRe}(\ket{\psi}\bra{\psi}_{AB})&=&{\rm Tr}_A\big\{(\varrho_A-\varrho_A^2)^{1/2}\big\}\nonumber\\
&:=&S_{\rm KD}^{\rm NRe}(\varrho_A)=S_{\rm KD}^{\rm NRe}(\varrho_B). 
\label{KD-nonreality entanglement as quantum entropy of the reduced density operator}
\end{eqnarray} 
\\ 
{\bf Proof}. See Appendix \ref{Proof of Proposition 3} for a proof.  

One can see that the right-hand side of Eq. (\ref{KD-nonreality entanglement as quantum entropy of the reduced density operator}), i.e., $S_{\rm KD}^{\rm NRe}(\varrho_{A(B)})$, quantifies the classical mixedness or quantum impurity of the quantum state of the subsystems $\varrho_{A(B)}={\rm Tr}_{B(A)}\ket{\psi}\bra{\psi}_{AB}$, i.e., it is vanishing for any pure state of the subsystems satisfying $\varrho_{A(B)}^2=\varrho_{A(B)}$ corresponding to a factorizable pure state $\ket{\psi}_{AB}=\ket{\psi_A}_A\ket{\psi_B}_B$, and it is maximized by the maximimally mixed state of the subsystems $\varrho_{A(B)}=\mathbb{I}/d$ corresponding to a maximally entangled pure state $\ket{\psi}_{AB}$. In the latter case, we have $E_{\rm KD}^{\rm NRe}(\ket{\psi}\bra{\psi}_{AB})=\sqrt{d-1}$ in accord with Proposition 1. Moreover, $S_{\rm KD}^{\rm NRe}(\varrho_{A(B)})$ is continuous and Schur concave function of $\varrho_{A(B)}$. It can thus be seen as a form of nonadditive quantum entropy of the state of the subsystems $\varrho_{A(B)}$. Hence, the pure state bipartite entanglement reveals itself in the minimum local uncertainty of the subsystems quantified by the specific form of nonadditive quantum entropy. It is remarkable that the nonadditive quantum entropy of the quantum state of the subsystems appears naturally as a certain extremal of the nonreality of the KD quasiprobability associated with the corresponding bipartite pure state.  

As a simple corollary, applying the Jensen inequality to Eq. (\ref{KD-nonreality entanglement as quantum entropy of the reduced density operator}), the normalized KD-nonreality entanglement for pure bipartite state is upper bounded  as follows 
\begin{eqnarray}
\widetilde{E}_{\rm KD}^{\rm NRe}(\ket{\psi}\bra{\psi}_{AB})&:=&\frac{E_{\rm KD}^{\rm NRe}(\ket{\psi}\bra{\psi}_{AB})}{\sqrt{d-1}}\nonumber\\
&\le&\sqrt{\frac{d}{d-1}(1-{\rm Tr}\{\varrho_A^2\})}:=C(\ket{\psi}\bra{\psi}_{AB}), 
\label{KD-nonreality entanglement for pure state is upper bounded by the entanglement concurrence}
\end{eqnarray}
where the right-hand side is just the entanglement concurrence of the bipartite pure state $\ket{\psi}_{AB}$ \cite{Wootters concurrence from entanglement of formation 1998,Wootters entanglement concurrence 2001}. As can be easily checked, for two qubits, the inequality becomes an equality, i.e., both yield $\widetilde{E}_{\rm KD}^{\rm NRe}(\ket{\psi}\bra{\psi}_{AB})=C(\ket{\psi}\bra{\psi}_{AB}) =2\sqrt{\lambda_1\lambda_2}$, where $\sqrt{\lambda_1}$ and $\sqrt{\lambda_2}$ are the Schmidt coefficients of $\ket{\psi}_{AB}$. For a two-qubit pure state, noting the relation between the concurrence $C$ and the entropy of entanglement $E_{\rm ent}$ that is given by the von Neumann entropy of the state of the subsystem, the KD-nonreality entanglement for a two-qubit pure state, which has the form of the nonadditive quantum entropy of Eq. (\ref{KD-nonreality entanglement as quantum entropy of the reduced density operator}), is related to  the entropy of entanglement as $E_{\rm ent}(\ket{\psi}\bra{\psi}_{AB})=g(C(\ket{\psi}\bra{\psi}_{AB}))=g(\widetilde{E}_{\rm KD}^{\rm NRe}(\ket{\psi}\bra{\psi}_{AB}))$, where $g(x)=h_2(\frac{1+\sqrt{1+x^2}}{2})$ and $h_2(x)=-x\log x-(1-x)\log(1-x)$ is the binary Shannon entropy \cite{Gour review}. 

Next notice that the definition in Eq. (\ref{KD-nonreality entanglement for pure bipartite state}) shows that the KD-nonreality pure state bipartite entanglement captures a certain extremal form of the noncommutativity between the bipartite state $\ket{\psi}_{AB}$ and the product PVM $\{\Pi_{x_A}\otimes\mathbb{I}_B\}$ describing local projective measurement. Below we obtain a lower bound in terms of the extremal trace-norm noncommutativity between the state and any local Hermitian operator or equivalently in terms of the extremal quantum standard deviation of any local Hermitian operator in the state of the subsystem as follows.\\ 
{\bf Proposition 3}. The KD-nonreality entanglement in a pure bipartite state $\ket{\psi}_{AB}$ on a finite-dimensional Hilbert space $\mathcal{H}_{AB}$ is lower bounded as
\begin{eqnarray}
\label{KD nonreality is lower bounded by a normalized trace-norm asymmetry}
E_{\rm KD}^{\rm NRe}(\ket{\psi}\bra{\psi}_{AB}])&\ge&\inf_{\{\Pi_{x_A}\}\in\mathcal{M}_{\rm r1PVM}(\mathcal{H}_A)}\sup_{X_A\in\mathbb{H}(\mathcal{H}_A|\{\Pi_{x_A}\})}\|[X_A\otimes\mathbb{I}_B,\ket{\psi}\bra{\psi}_{AB}]\|_1/2\|X_A\|_{\infty}\\
\label{KD nonreality entanglement is lower bounded by normalized local variance}
&=&\inf_{\{\Pi_{x_A}\}\in\mathcal{M}_{\rm r1PVM}(\mathcal{H}_A)}\sup_{X_A\in\mathbb{H}(\mathcal{H}_A|\{\Pi_{x_A}\})}\Delta_{X_A}(\varrho_A)/\|X_A\|_{\infty}, 
\end{eqnarray}
where $\mathbb{H}(\mathcal{H}_A|\{\Pi_{x_A}\})$ is the convex set of all Hermitian operators $X_A$ acting on the Hilbert space $\mathcal{H}_A$ whose complete set of eigenprojectors is $\{\Pi_{x_A}\}$, and $\|O\|_{\infty}$ denotes the operator norm of $O$ \cite{Watrous book quantum information theory}. Moreover, the lower bound depends only on the Schmidt coefficients of $\ket{\psi}_{AB}$ or the eigenvalues of $\varrho_A$. \\
{\bf Proof}. See Appendix \ref{Proofs of Proposition 2} for a proof. 

The above proposition shows that nonvanishing lower bound in Eqs. (\ref{KD nonreality is lower bounded by a normalized trace-norm asymmetry}) or (\ref{KD nonreality entanglement is lower bounded by normalized local variance}) can be seen as a sufficient condition for detecting the bipartite entanglement in the state $\ket{\psi}_{AB}$. Notice that the term $\|[X_A\otimes\mathbb{I}_B,\ket{\psi}\bra{\psi}_{AB}]]\|_1/2$ in the lower bound in Eq. (\ref{KD nonreality is lower bounded by a normalized trace-norm asymmetry}) is just the trace-norm asymmetry in the state $\ket{\psi}_{AB}$ relative to the grup of translation unitaries generated by the product Hermitian operator $X_A\otimes\mathbb{I}_B$  \cite{Marvian - Spekkens speakable and unspeakable coherence}. Hence, we have linked the bipartite entanglement to the extremal of the local translational asymmetry, another important form of quantum resource for various quantum protocols, via the nonreality of the KD quasiprobability. Finally, Eq. (\ref{KD nonreality entanglement is lower bounded by normalized local variance}) shows that the KD-nonreality bipartite entanglement in the pure state $\ket{\psi}_{AB}$ is lower bounded by the extremal of the uncertainty of any local Hermitian operator in the state of the subsystem as quantified by the quantum standard deviation.  

Let us proceed to discuss an operational and geometrical interpretation of the KD-nonreality pure state entanglement. First, we have the following result. \\
{\bf Proposition 4}. Let $\varrho_{AB;x_A}$ denote the state after a nonselective binary measurement described by the PVM $\{\Pi_{x_A}\otimes\mathbb{I}_B,\mathbb{I}-(\Pi_{x_A}\otimes\mathbb{I}_B)\}$ over the pure state $\ket{\psi}_{AB}$. Then, we have
\begin{eqnarray}
E_{\rm KD}^{\rm NRe}(\ket{\psi}\bra{\psi}_{AB})&=&\inf_{\{\Pi_{x_A}\}\in\mathcal{M}_{\rm r1PVM}(\mathcal{H}_A)}\frac{1}{2}\sum_{x_A}\|\ket{\psi}\bra{\psi}_{AB}-\varrho_{AB;x_A}\|_1
\label{KD-nonreality entanglement as trace-norm of local measurement disturbance}\\
&\ge&\inf_{\{\Pi_{x_A}\}\in\mathcal{M}_{\rm r1PVM}(\mathcal{H}_A)}\frac{1}{2}\sum_{x_A}\|\varrho_A-\varrho_{A;x_A}\|_1.
\label{KD-nonreality entanglement is upper bounded by trace-norm of local nonselective measurement disturbance}
\end{eqnarray}
Here, $\varrho_{A;x_A}$ is the state of the subsystem A after the nonselective binary measurement described by the PVM $\{\Pi_{x_A},\mathbb{I}_A-\Pi_{x_A}\}$ over the state $\varrho_A={\rm Tr}_B\{\ket{\psi}\bra{\psi}_{AB}\}$. 
\\
{\bf Proof}. See Appendix \ref{proof of proposition 4} for a proof.\\
Equation (\ref{KD-nonreality entanglement as trace-norm of local measurement disturbance}) suggests that the KD-nonreality pure state bipartite entanglement can be seen as to capture the minimum total state disturbance due to any local binary projective measurement over the global state. Moreover, Eq. (\ref{KD-nonreality entanglement is upper bounded by trace-norm of local nonselective measurement disturbance}) shows that it is lower bounded by the minimum disturbance due to any nonselective local binary measurement on the subsystem.  

Finally, we define a measure of bipartite entanglement for a generic state using that for the pure state and its convex roof extension as follows.\\
{\bf Definition 2}. The KD-nonreality bipartite entanglement in a generic state $\varrho_{AB}$ is defined as:
\begin{eqnarray}
E_{\rm KD}^{\rm NRe}(\varrho_{AB})&=&\inf_{\substack{\{p_k,\ket{\psi^k}\bra{\psi^k}_{AB}\}\\ \varrho_{AB}=\sum_kp_k\ket{\psi^k}\bra{\psi^k}_{AB}}}\sum_k p_kE_{\rm KD}^{\rm NRe}(\ket{\psi^k}\bra{\psi^k}_{AB})\nonumber\\
&=&\inf_{\substack{\{p_k,\ket{\psi^k}\bra{\psi^k}_{AB}\}\\ \varrho_{AB}=\sum_kp_k\ket{\psi^k}\bra{\psi^k}_{AB}}}\sum_k p_k{\rm Tr}_A\{(\varrho_A^k-(\varrho_A^k)^2)^{1/2}\},
\label{KD-nonreality entanglement for general state} 
\end{eqnarray}
Here, the infimum is computed with respect to all pure states decomposition of the bipartite state $\varrho_{AB}$, i.e., $\varrho_{AB}=\sum_kp_k\ket{\psi^k}\bra{\psi^k}_{AB}$, $p_k\ge 0$, $\sum_kp_k=1$, $\varrho_A^k={\rm Tr}_B\{\ket{\psi^k}\bra{\psi^k}_{AB}\}$, and the second equality is due Eq. (\ref{KD-nonreality entanglement as quantum entropy of the reduced density operator}).

From the above definition, and Theorem 1, we have the following theorem.\\
{\bf Theorem 2}. $E_{\rm KD}^{\rm NRe}(\varrho_{AB})$ defined in Eq. (\ref{KD-nonreality entanglement for general state}) is a convex and monotonic measure of entanglement. 

We thus call $E_{\rm KD}^{\rm NRe}(\varrho_{AB})$ defined in Eq. (\ref{KD-nonreality entanglement for general state}) the KD-nonreality bipartite entanglement in the state $\varrho_{AB}$. From Eqs. (\ref{KD-nonreality entanglement for pure state is upper bounded by the entanglement concurrence}) and (\ref{KD-nonreality entanglement for general state}), one concludes that the KD-nonreality entanglement normalized by its maximum value is upper bounded by the concurrence of formation. The latter is defined as the convex roof extension of the concurrence for pure state. Due to the equality between the normalized KD-nonreality entanglement and the concurrence for a two-qubit pure state, the normalized KD-nonreality entanglement in a generic two-qubit state is also equal to the concurrence of formation.    

Moreover, from the definition of Eq. (\ref{KD-nonreality entanglement for general state}), Eq. (\ref{KD-nonreality entanglement as quantum entropy of the reduced density operator}) of Proposition 2, and Eq. (\ref{KD nonreality entanglement is lower bounded by normalized local variance}) of Proposition 3, we have the following result. \\
{\bf Proposition 5}. The KD-nonreality bipartite entanglement in a generic state $\varrho_{AB}$ is lower and upper bounded as:
\begin{eqnarray}
\inf_{\{\Pi_{x_A}\}\in\mathcal{B}_{\rm o}(\mathcal{H}_A)}\sup_{X_A\in\mathbb{H}(\mathcal{H}_A|\{\Pi_{x_A}\})}\|[X_A\otimes\mathbb{I}_B,\varrho_{AB}]\|_1/2\|X_A\|_{\infty}\nonumber\\
\le E_{\rm KD}^{\rm NRe}(\varrho_{AB})\le S_{\rm KD}^{\rm NRe}(\varrho_A), 
\label{upper and lower bounds for the KD-nonreality entanglement for generic state}
\end{eqnarray}
where $\varrho_A={\rm Tr}_B\{\varrho_{AB}\}$. 
\\
{\bf Proof}. See Appendix \ref{Proof of Theorem 2} for a proof.\\
Again, the lower bound takes the form of an extremum of the trace-norm asymmetry in the state $\varrho_{AB}$ relative to the translation unitaries generated by local Hermitian operator of the form $X_A\otimes\mathbb{I}_B$. Notice that the upper bound takes its maximum when $\varrho_A=\mathbb{I}/d$, which corresponds to the maximally entangled pure bipartite state $\ket{\psi}_{AB}$. Moreover, in this case, the second inequality in Eq. (\ref{upper and lower bounds for the KD-nonreality entanglement for generic state}) becomes an equality. Next, recall that when the bipartite state is pure, $\varrho_{AB}=\ket{\psi}\bra{\psi}_{AB}$, the lower bound is given by Eq. (\ref{KD nonreality entanglement is lower bounded by normalized local variance}) in terms of a certain extremum of the variance of Hermitian operator $X_A$ over $\varrho_A$ of subsystem A. In this sense, Eq. (\ref{upper and lower bounds for the KD-nonreality entanglement for generic state}) shows that the KD-nonreality entanglement for the generic state is lower and upper bounded by two different forms of quantum uncertainty in the subsystem quantified, respectively, by an extremal trace-norm asymmetry and a nonadditive quantum entropy. Note however that, unlike the case of pure state, the lower and upper bounds in Eq. (\ref{upper and lower bounds for the KD-nonreality entanglement for generic state}) are now in general no longer symmetric with respect to the exchange of A and B. Hence, the lower and upper bounds can be further strengthened, respectively by choosing the larger and smaller bounds, obtained by exchanging A and B. 

\section{Estimation of the KD bipartite entanglement using weak value measurement} 

We briefly sketch the estimation of the KD-nonreality bipartite entanglement, without full state tomography, based on weak value estimation as follows. First, the KD quasiprobability defined in Eq. (\ref{KD quasiprobability with one variable for the first orthonormal product basis}) can be expressed as ${\rm Pr}_{\rm KD}(x_A;y_{AB}|\varrho_{AB})=\Pi_{x_A}^{\rm w}(y_{AB}|\varrho_{AB}){\rm Pr}(y_{AB}|\varrho_{AB})$, where $\Pi_{x_A}^{\rm w}(y_{AB}|\varrho_{AB}):=\frac{{\rm Tr}\{\Pi_{y_{AB}}(\Pi_{x_A}\otimes\mathbb{I}_B)\varrho_{AB}\}}{{\rm Tr}\{\Pi_{y_{AB}}\varrho_{AB}\}}$, is the weak value of an element of the local PVM $(\Pi_{x_A}\otimes\mathbb{I}_B)$ with the preselected state $\varrho_{AB}$ and the postselected pure state $\ket{y_{AB}}$, and ${\rm Pr}(y_{AB}|\varrho_{AB})={\rm Tr}\{\Pi_{y_{AB}}\varrho_{AB}\}$ is just the probability to get $y_{AB}$ in a strong projective measurement \cite{Aharonov weak value,Aharonov-Daniel book,Wiseman weak value}. The weak value $\Pi_{x_A}^{\rm w}(y_{AB}|\varrho_{AB})$ can be estimated directly in experiments using various schemes \cite{Aharonov weak value,Aharonov-Daniel book,Wiseman weak value,Lundeen complex weak value,Jozsa complex weak value,Johansen quantum state from successive projective measurement,Vallone strong measurement to reconstruct quantum wave function,Cohen estimating of weak value with strong measurements,Hernandez-Gomez experimental observation of TBMH negativity,Wagner measuring weak values and KD quasiprobability,Chiribella dimension independent weak value estimation independent}. Noting this, the KD-nonreality bipartite entanglement of an unknown quantum state $\varrho_{AB}$ can thus be, in principle, estimated by combining such an estimation of the weak value and maximization over all possible orthonormal bases of $\mathcal{H}_{AB}$ and minimization over all orthonormal bases of $\mathcal{H}_A$. These optimizations can be carried out classically. In this variational quantum algorithm, we assume a unitary circuit that rotates the standard orthonormal basis into any orthonormal bases of $\mathcal{H}_{AB}$ and $\mathcal{H}_A$. Note that for a mixed bipartite state, we also need a pre-processing quantum circuit which decomposes the state into all possible mixes of pure states. To avoid the minimization over all possible pure states decomposition, we can instead experimentally estimate the lower and upper bounds given in Eq. (\ref{upper and lower bounds for the KD-nonreality entanglement for generic state}). The trace-norm asymmetry in the lower bound can be estimated again using the weak value measurement and classical optimization by expressing it in terms of extremum imaginary part of weak value as shown in Refs. \cite{Agung trace-norm asymmetry from nonreal weak value,Agung asymmetry from nonreal weak value}. The upper bound can also be obtained using local projective measurement and classical optimization following the method of proof of Proposition 2 in Appendix \ref{Proof of Proposition 3}. 

The above results reveal a intriguing connection between entanglement and generalized quantum contextuality \cite{Spekkens generalized quantum contextuality}. First, it has been shown that when the weak value is nonreal, its estimation using weak measurement with postselection cannot be modelled by any noncontextual ontological (hidden variable) model \cite{Kunjwal contextuality of non-real weak value}. This implies that a nonvanishing KD-nonreality entanglement is sufficient to prove general quantum contextuality. This can be seen by noting the fact that when the KD-nonreality entanglement in a state $\varrho_{AB}$ is nonvanishing, there must be a local rank-1 PVM $\{\Pi_{x_A}\}$ and an orthonormal basis $\{\ket{y_{AB}}\}$ so that the weak value $\Pi_{x_A}^{\rm w}(y_{AB}|\varrho_{AB})$ of $\Pi_{x_A}\otimes\mathbb{I}_B$ with the preselected state $\varrho_{AB}$ and postselected state $\ket{y_{AB}}$ must be nonreal for some $x_A$ and $y_{AB}$, so that  its estimation based on weak measurement with postselection cannot be modelled by any noncontextual ontological model. Conversely, the above observation also shows that the KD-nonreality entanglement can be seen as a specific extremal form of generalized quantum contextuality via weak measurement with postselection.   

\section{Conclusion}

We have developed a bipartite entanglement measure, called KD-nonreality bipartite entanglement, using the nonreal values of the KD quasiprobability over a pair of orthonormal bases one of which is a product, and suitable optimizations over all possible choices of these bases. For pure states, we obtained an analytical expression for the bipartite entanglement measure in terms of a specific form of nonadditive quantum entropy of the state of the subsystems. The entanglement measure for generic states was then constructed using convex roof extension. The normalized form of the entanglement measure provides a lower bound to the concurrence of formation and they are equal for two-qubit state. We also derived upper and lower bounds in terms of different forms of uncertainty in the subsystems, and offered geometrical interpretation in terms of the trace distance between the state of interest and the final state after a certain nonselective local binary measurement. We showed that the KD-nonreality bipartite entanglement can in principle be estimated via weak value measurement using various schemes, combined with classical optimization. This reveals an interesting link between entanglement and quantum contextuality via the nonreal values of the KD quasiprobability. It is interesting to apply a similar logic to quantify entanglement as an extremal value of the trace-norm translational asymmetry which has recently also been related to the imaginary part of the weak value \cite{Agung asymmetry from nonreal weak value,Agung trace-norm asymmetry from nonreal weak value}. Our results not only solve a long standing intriguing question about the relation between entanglement and the anomalous values of quasiprobability, but might open a fresh avenue of applications of entanglement and nonclassical values of Kirkwood-Dirac quasiprobability in different areas of quantum science. 

\begin{acknowledgments}  
This work is partly funded by the Institute for Research and Community Service, Bandung Institute of Technology with the grant number: 2971/IT1.B07.1/TA.00/2021. It is also in part supported by the Indonesia Ministry of Research, Technology, and Higher Education with the grant number: 187/E5/PG.02.00.PT/2022 and 2/E1/KP.PTNBH/2019. The Author would like to thank the anonymous Referees for constructive criticism and suggestions, and Joel Federicko Sumbowo for useful discussion. 
\end{acknowledgments} 

\appendix

\section{Proof of properties 1 - 2\label{Proof of properties ENR1-ENR3}}

We prove Properties 1 and 2 stated in the main text. \\
{\it Proof of Property 1}. First, suppose that the bipartite pure state is factorizable as: $\ket{\psi}_{AB}=\ket{\psi_A}_A\ket{\psi_B}_B$. Then, we can choose an orthonormal basis $\{\ket{x_A}\}$ of $\mathcal{H}_A$ in the definition of $E_{\rm KD}^{\rm NRe}(\ket{\psi}\bra{\psi}_{AB})$ in Eq. (\ref{KD-nonreality entanglement for pure bipartite state}) such that one of its elements is given by $\ket{\psi_A}$. In this case, one evidently has $[(\Pi_{x_A}\otimes\mathbb{I}_B),\ket{\psi_A}\bra{\psi_A}\otimes\ket{\psi_B}\bra{\psi_B}]=0$ for all $x_A$, so that we have $E_{\rm KD}^{\rm NRe}(\ket{\psi}\bra{\psi}_{AB})=0$ as per the definition in Eq. (\ref{KD-nonreality entanglement for pure bipartite state}). Conversely, suppose $E_{\rm KD}^{\rm NRe}(\ket{\psi}\bra{\psi}_{AB})=0$. Then, from the definition in Eq. (\ref{KD-nonreality entanglement for pure bipartite state}), there must be an orthonormal basis $\{\ket{x_A}\}$ of $\mathcal{H}_A$ such that $\braket{y_{AB}|[(\Pi_{x_A}\otimes\mathbb{I}_B),\ket{\psi}\bra{\psi}_{AB}]|y_{AB}}=0$ for all $x_A$ and for all orthonormal bases $\{\ket{y_{AB}}\}$ of $\mathcal{H}_{AB}$. This can only be true if for all the elements of the orthonormal basis $\{\ket{x_A}\}$ we have $[(\Pi_{x_A}\otimes\mathbb{I}_B),\ket{\psi}\bra{\psi}_{AB}]=0$. Hence, $\ket{\psi}_{AB}$ must take the form of $\ket{\psi}_{AB}=\ket{x_A}\ket{\psi_B}$ for some $x_A$, as can be seen by using the Schmidt decomposition theorem for bipartite pure state. \qed
\\
{\it Proof of property 2}. This comes directly from the definition as follows: 
\begin{eqnarray}
&&E_{\rm KD}^{\rm NRe}((U_A\otimes U_B)\ket{\psi}\bra{\psi}_{AB}(U_A^{\dagger}\otimes U_B^{\dagger}))\nonumber\\
&=&\inf_{\{\ket{x_A}\}\in\mathcal{B}_{\rm o}(\mathcal{H}_A)}\sum_{x_A}\sup_{\{\ket{y_{AB}}\}\in\mathcal{B}_{\rm o}(\mathcal{H}_{AB})}\sum_{y_{AB}}\big|{\rm Im}(\langle y_{AB}|(U_A\otimes U_B)(U_A^{\dagger}\otimes U_B^{\dagger})\nonumber\\
&&\hspace{20mm}\cdot\hspace{1mm}(\Pi_{x_A}\otimes\mathbb{I}_B)(U_A\otimes U_B)\ket{\psi}\bra{\psi}_{AB}(U_A^{\dagger}\otimes U_B^{\dagger})|y_{AB}\rangle)\big|\nonumber\\
&=&\inf_{\{\ket{x_A}\}\in\mathcal{B}_{\rm o}(\mathcal{H}_A)}\sum_{x_A^{U_A}}\sup_{\{\ket{y_{AB}}\}\in\mathcal{B}_{\rm o}(\mathcal{H}_{AB})}\sum_{y_{AB}^{U_AU_B}}\big|{\rm Im}(\braket{y_{AB}^{U_AU_B}|(\Pi_{x_A}^{U_A}\otimes\mathbb{I}_B)(|\psi}\braket{\psi|_{AB})|y_{AB}^{U_AU_B}})\big|\nonumber\\
&=&\inf_{\{\ket{x_A^{U_A}}\}\in\mathcal{B}_{\rm o}(\mathcal{H}_A)}\sum_{x_A^{U_A}}\sup_{\{\ket{y_{AB}^{U_AU_B}}\}\in\mathcal{B}_{\rm o}(\mathcal{H}_{AB})}\sum_{y_{AB}^{U_AU_B}}\big|{\rm Im}(\braket{y_{AB}^{U_AU_B}|(\Pi_{x_A}^{U_A}\otimes\mathbb{I}_B)(|\psi}\braket{\psi|_{AB})|y_{AB}^{U_AU_B}})\big|\nonumber\\
&=&E_{\rm KD}^{\rm NRe}(\ket{\psi}\bra{\psi}_{AB}).
\label{invariant under local unitary}
\end{eqnarray}
Here, we have inserted an identity  $(U_A\otimes U_B)(U_A^{\dagger}\otimes U_B^{\dagger})=\mathbb{I}$ in the second line. To get the third line, we have defined a new orthonormal basis: $\{\ket{y_{AB}^{U_AU_B}}=(U_A^{\dagger}\otimes U_B^{\dagger})\ket{y_{AB}}\}$, and identified a new rank-1 PVM basis: $\{\Pi_{x_A}^{U_A}\}=\{U_A^{\dagger}\Pi_{x_A}U_A\}$ corresponding to a new orthonormal basis $\{\ket{x_A^{U_A}}\}=\{U_A^{\dagger}\ket{x_A}\}$. Noting that the above local transformations of bases do not change the sets $\mathcal{B}_{\rm o}(\mathcal{H}_{AB})$ and $\mathcal{B}_{\rm o}(\mathcal{H}_A)$ over which we perform, respectively, the maximization and the minimization in the definition of the KD-nonreality pure state bipartite entanglement in Eq. (\ref{KD-nonreality entanglement for pure bipartite state}), we thus have  $\sup_{\{\ket{y_{AB}}\}\in\mathcal{B}_{\rm o}(\mathcal{H}_{AB})}(\cdot)=\sup_{\{\ket{y_{AB}^{U_AU_B}}\}\in\mathcal{B}_{\rm o}(\mathcal{H}_{AB})}(\cdot)$ and $\inf_{\{\ket{x_A}\}\in\mathcal{B}_{\rm o}(\mathcal{H}_A)}(\cdot)=\inf_{\{\ket{x_A^{U_A}}\}\in\mathcal{B}_{\rm o}(\mathcal{H}_A)}(\cdot)$. This observation gives the last equality in Eq. (\ref{invariant under local unitary}). \qed 

\section{Proof of Proposition 1 \label{Proof of first part of Proposition 1}}

First, we state the following lemma. \\
{\bf Lemma 1}. Given a normal operator $O$ on a Hilbert space $\mathcal{H}$, the trace-norm or the Schatten $p=1$ norm of $O$ can be expressed as a variational problem as
\begin{eqnarray}
\|O\|_1:={\rm Tr}\{|O|\}=\sup_{\{\ket{y}\}\in\mathcal{B}_{\rm o}(\mathcal{H})}\sum_y|\braket{y|O|y}|,
\label{Lemma on the variational expression on the trace-norm}
\end{eqnarray}
where $|O|:=\sqrt{OO^{\dagger}}$, and the supremum is taken over the set $\mathcal{B}_{\rm o}(\mathcal{H})$ of all orthonormal bases of $\mathcal{H}$. 
 \\
{\bf Proof}. The proof for the skew Hermitian operators is given in Ref. \cite{Agung trace-norm asymmetry from nonreal weak value} (see Proposition 1 of that paper). The same steps of proof apply for all normal operators. First, since $O$ is a normal operator, it has a spectral decomposition $O=\sum_io_i\ket{o_i}\bra{o_i}$, where $\{o_i\}$ is the set of in general complex eigenvalues of $O$ with the corresponding orthonormal set of eigenvectors $\{\ket{o_i}\}$. We thus have, upon inserting this into the right-hand side of Eq. (\ref{Lemma on the variational expression on the trace-norm}),  
\begin{eqnarray}
\label{trace norm from maximum absolute expectation value over all bases step 1}
\sup_{\{\ket{y}\}\in\mathcal{B}_{\rm o}(\mathcal{H})}\sum_y|\braket{y|O|y}|&=&\sup_{\{\ket{y}\}\in\mathcal{B}_o(\mathcal{H})}\sum_{y}\big|\sum_io_i\braket{y|o_i}\braket{o_i|y}\big|\nonumber\\
\label{trace norm from maximum absolute expectation value over all bases step 2}
&=&\sum_{y_*}\big|\sum_io_i\braket{y_*|o_i}\braket{o_i|y_*}\big|\\
\label{trace norm from maximum absolute expectation value over all bases step 3}
&\le&\sum_i|o_i|\sum_{y_*}|\braket{y_*|o_i}|^2\\
\label{trace norm from maximum absolute expectation value over all bases step 3.5}
&=&\sum_i|o_i|={\rm Tr}\{|O|\}=\|O\|_1.
\end{eqnarray}
Here, $\{\ket{y_*}\}$ in Eq. (\ref{trace norm from maximum absolute expectation value over all bases step 2}) is an orthonormal basis which attaines the supremum, the inequality in Eq. (\ref{trace norm from maximum absolute expectation value over all bases step 3}) is due to the triangle inequality, and we have used the completeness relation for $\{\ket{y_*}\}$ to obtain Eq. (\ref{trace norm from maximum absolute expectation value over all bases step 3.5}). On the other hand, one can see in Eq. (\ref{trace norm from maximum absolute expectation value over all bases step 2}) that the equality in Eq. (\ref{trace norm from maximum absolute expectation value over all bases step 3}), i.e., the upper bound, can always be attained by choosing $\{\ket{y_*}\}=\{\ket{o_i}\}$ so that we get Eq. (\ref{Lemma on the variational expression on the trace-norm}). \qed. 
\\ \\
Using Eq. (\ref{Lemma on the variational expression on the trace-norm}) of Lemma 1, noting that $\Pi_{x_A}\otimes \mathbb{I}_B$ is Hermitian for all $x_A$ so that $[\ket{\psi}\bra{\psi}_{AB},(\Pi_{x_A}\otimes\mathbb{I}_B)]$ is skew Hermitian, we thus obtain from Eq. (\ref{KD-nonreality entanglement for pure bipartite state}),
\begin{eqnarray}
&&E_{\rm KD}^{\rm NRe}(\ket{\psi}\bra{\psi}_{AB})\nonumber\\
&=&\inf_{\{\Pi_{x_A}\}\in\mathcal{M}_{\rm r1PVM}(\mathcal{H}_A)}\sum_{x_A}\sup_{\{\ket{y_{AB}}\in\mathcal{B}_{\rm o}(\mathcal{H}_{AB})\}}\sum_{y_{AB}}\frac{1}{2}\big|\braket{y_{AB}|[(\Pi_{x_A}\otimes \mathbb{I}_B),\ket{\psi}\bra{\psi}_{AB}]|y_{AB}}\big|\nonumber\\
&=&\inf_{\{\Pi_{x_A}\}\in\mathcal{M}_{\rm r1PVM}(\mathcal{H}_A)}\sum_{x_A}\|[\ket{\psi}\bra{\psi}_{AB},(\Pi_{x_A}\otimes\mathbb{I}_B)]\|_1/2,
\label{proof of Proposition 1 step 3} 
\end{eqnarray}
where $\mathcal{M}_{\rm r1PVM}(\mathcal{H}_A)$ is the set of all rank-1 PVMs on $\mathcal{H}_A$ associated with the set $\mathcal{B}_{\rm o}(\mathcal{H}_A)$ of orthonormal bases of $\mathcal{H}_A$. The right-hand side can be seen as the infimum of the total noncommutativity, as quantified by the trace norm,  between the bipartite pure state $\varrho_{AB}=\ket{\psi}\bra{\psi}_{AB}$ and any PVM of the form $\{\Pi_{x_A}\otimes\mathbb{I}_B\}$ describing local measurement on  subsystem A while leaving subsystem B untouched. Note further that $\|[\ket{\psi}\bra{\psi}_{AB},(\Pi_{x_A}\otimes\mathbb{I}_B)]\|_1/2$ can be interpreted as the trace-norm asymmetry in the state $\ket{\psi}\bra{\psi}_{AB}$ relative to the group of translation unitaries generated by the product Hermitian operator $\Pi_{x_A}\otimes\mathbb{I}_B$ \cite{Marvian - Spekkens speakable and unspeakable coherence}. 

Next, we use the known equality between the trace-norm noncommutativity between any Hermitian operator $O$ and pure state $\ket{\psi}$, and the quantum standard deviation of $O$ in $\ket{\psi}$ \cite{Marvian - Spekkens speakable and unspeakable coherence}, i.e., 
\begin{eqnarray}
\|[O,\ket{\psi}\bra{\psi}]\|_1/2&=&({\rm Tr}\{O^2\ket{\psi}\bra{\psi}\}-{\rm Tr}\{O\ket{\psi}\bra{\psi}\}^2)^{1/2 }\nonumber\\
&:=&\Delta_O(\ket{\psi}\bra{\psi}), 
\label{trace-norm asymmetry is equal to the quantum standard deviation}
\end{eqnarray}
to obtain   
\begin{eqnarray}
&&E_{\rm KD}^{\rm NRe}(\ket{\psi}\bra{\psi}_{AB})\nonumber\\
&=&\inf_{\{\Pi_{x_A}\}\in\mathcal{M}_{\rm r1PVM}(\mathcal{H}_A)}\sum_{x_A}\|[\ket{\psi}\bra{\psi}_{AB},(\Pi_{x_A}\otimes\mathbb{I}_B)]\|_1/2\nonumber\\
&=&\inf_{\{\Pi_{x_A}\}\in\mathcal{M}_{\rm r1PVM}(\mathcal{H}_A)}\sum_{x_A}\Delta_{(\Pi_{x_A}\otimes\mathbb{I}_B)}(\ket{\psi}\bra{\psi}_{AB})\nonumber\\
&=&\inf_{\{\Pi_{x_A}\}\in\mathcal{M}_{\rm r1PVM}(\mathcal{H}_A)}\sum_{x_A}({\rm Tr}\{(\Pi_{x_A}\otimes\mathbb{I}_B)^2\ket{\psi}\bra{\psi}_{AB}\}-{\rm Tr}\{(\Pi_{x_A}\otimes\mathbb{I}_B)\ket{\psi}\bra{\psi}_{AB}\}^2)^{1/2}\nonumber\\
&=&\inf_{\{\Pi_{x_A}\}\in\mathcal{M}_{\rm r1PVM}(\mathcal{H}_A)}\sum_{x_A}({\rm Tr}\{\Pi_{x_A}^2\varrho_A\}-{\rm Tr}\{\Pi_{x_A}\varrho_A\}^2)^{1/2}\nonumber\\ 
&=&\inf_{\{\Pi_{x_A}\}\in\mathcal{M}_{\rm r1PVM}(\mathcal{H}_A)}\sum_{x_A}\Delta_{\Pi_{x_A}}(\varrho_A)\nonumber\\
&=&\inf_{\{\Pi_{x_A}\}\in\mathcal{M}_{\rm r1PVM}(\mathcal{H}_A)}\sum_{x_A}\big({\rm Pr}(x_A|\varrho_A)-{\rm Pr}(x_A|\varrho_A)^2\big)^{1/2}, 
\label{proof of Proposition 1 step 4} 
\end{eqnarray}
where $\varrho_A={\rm Tr}_B\{\ket{\psi}\bra{\psi}_{AB}\}$, and in the last line we have used ${\rm Tr}_A\{\Pi_{x_A}^2\varrho_A\}={\rm Tr}_A\{\Pi_{x_A}\varrho_A\}={\rm Pr}(x_A|\varrho_A)$. Hence, we have proved Eqs. (\ref{KD entanglement versus local quantum uncertainty step 1})-(\ref{KD entanglement versus local quantum uncertainty step 3}) of Proposition 1. 

Finally, let us show that the maximum value of $E_{\rm KD}^{\rm NRe}(\ket{\psi}\bra{\psi}_{AB})$ is obtained when the state $\ket{\psi}_{AB}$ is maximally entangled giving Eq. (\ref{KD entanglement versus local quantum uncertainty step 4}). First, from Eq. (\ref{proof of Proposition 1 step 4}), upon applying the Jensen inequality, we have
\begin{eqnarray}
&&E_{\rm KD}^{\rm NRe}(\ket{\psi}\bra{\psi}_{AB})\nonumber\\
&=&\inf_{\{\Pi_{x_A}\}\in\mathcal{M}_{\rm r1PVM}(\mathcal{H}_A)}\sum_{x_A}\big({\rm Pr}(x_A|\varrho_A)-{\rm Pr}(x_A|\varrho_A)^2\big)^{1/2}\nonumber\\
&\le&\inf_{\{\Pi_{x_A}\}\in\mathcal{M}_{\rm r1PVM}(\mathcal{H}_A)}d ^{1/2}\big(1-\sum_{x_A}{\rm Pr}(x_A|\varrho_A)^2\big)^{1/2}. 
\label{proof of Proposition 1 step 5} 
\end{eqnarray}
The equality is obtained when the probability ${\rm Pr}(x_A|\varrho_A)$ is uniform, so that, imposing normalization, we have ${\rm Pr}(x_A|\varrho_A)=1/d$, where $d$ is the Schmidt rank, namely, the state of subsystem A is given by a totally mixed state $\varrho_A=\mathbb{I}_A/d$. This is the case when the bipartite pure state has the Schmidt decomposition: $\ket{\psi^{\rm me}}_{AB}=\sum_k\frac{1}{\sqrt{d }}\ket{\mu^k}_A\ket{\nu^k}_B$, where $\{\ket{\mu^k}_A\}$ and $\{\ket{\nu^k}_B\}$ are subsets of orthonormal bases of $\mathcal{H}_A$ and $\mathcal{H}_B$, respectively. Hence, the state is maximally entangled. Since the maximally mixed state of subsystem A returns a uniform probability distribution for all measurements described by any rank-1 PVM $\{\Pi_{x_A}\}$, we have 
\begin{eqnarray}
&&E_{\rm KD}^{\rm NRe}(\ket{\psi^{\rm me}}\bra{\psi^{\rm me}}_{AB})\nonumber\\
&=&\inf_{\{\Pi_{x_A}\}\in\mathcal{M}_{\rm r1PVM}(\mathcal{H}_A)}\sum_{x_A}\big(1/d -1/d ^2\big)^{1/2}\nonumber\\
&=&\sqrt{d -1}. 
\label{proof of Proposition 1 step 6} 
\end{eqnarray}
\qed

\section{Proof of Proposition 2\label{Proof of Proposition 3}}

We show that the infimum in Eq. (\ref{proof of Proposition 1 step 4}) is attained when the rank-1 PVM $\{\Pi_{x_A}\}\in\mathcal{M}_{\rm r1PVM}(\mathcal{H}_A)$ is given by the eigenprojectors of the reduced state $\varrho_A={\rm Tr}_B\{\ket{\psi}\bra{\psi}_{AB}\}$ of subsystem A. To see this, first, assume that $\varrho_A$ has the spectral decomposition: $\varrho_A=\sum_j\lambda_j(\varrho_A)\Pi_{\lambda_j(\varrho_A)}$, where $\{\Pi_{\lambda_j(\varrho_A)}=\ket{\lambda_j(\varrho_A)}\bra{\lambda_j(\varrho_A)}\}$ is the complete set of the eigenprojectors of $\varrho_A$, and $\{\lambda_j(\varrho_A)\}_{j=1}^d$, with $\lambda_j(\varrho_A)> 0$, $\sum_j\lambda_j(\varrho_A)=1$, is the set of associated nonvanishing eigenvalues. Hence, $\{\lambda_j(\varrho_A)^{1/2}\}$ is just the set of Schmidt coefficients of the original pure bipartite state $\ket{\psi}_{AB}$. We thus have ${\rm Pr}(x_A|\varrho_A)={\rm Tr}\{\Pi_{x_A}\varrho_A\}=\sum_j\lambda_j(\varrho_A){\rm Tr}\{\Pi_{\lambda_j(\varrho_A)}\Pi_{x_A}\}$. Using this and the Jensen inequality, we obtain
\begin{eqnarray}
&&\sum_{x_A}\big({\rm Pr}(x_A|\varrho_A)-{\rm Pr}(x_A|\varrho_A)^2\big)^{1/2}\nonumber\\
&=&\sum_{x_A}\big(\sum_j\lambda_j(\varrho_A){\rm Tr}\{\Pi_{\lambda_j(\varrho_A)}\Pi_{x_A}\}-\big(\sum_k\lambda_k(\varrho_A){\rm Tr}\{\Pi_{\lambda_k(\varrho_A)}\Pi_{x_A}\}\big)^2\big)^{1/2}\nonumber\\
&\ge&\sum_{x_A}\big(\sum_j\lambda_j(\varrho_A){\rm Tr}\{\Pi_{\lambda_j(\varrho_A)}\Pi_{x_A}\}-\sum_k\lambda_k(\varrho_A)^2{\rm Tr}\{\Pi_{\lambda_k(\varrho_A)}\Pi_{x_A}\}\big)^{1/2}\nonumber\\
&=&\sum_{x_A}\big(\sum_j(\lambda_j(\varrho_A)-\lambda_j(\varrho_A)^2){\rm Tr}\{\Pi_{\lambda_j(\varrho_A)}\Pi_{x_A}\}\big)^{1/2}\nonumber\\
&\ge&\sum_{x_A}\sum_j\big(\lambda_j(\varrho_A)-\lambda_j(\varrho_A)^2\big)^{1/2}{\rm Tr}\{\Pi_{\lambda_j(\varrho_A)}\Pi_{x_A}\}\nonumber\\
&=&\sum_j\big(\lambda_j(\varrho_A)-\lambda_j(\varrho_A)^2\big)^{1/2}. 
\label{infimum of the KD-nonreality entanglement 1}
\end{eqnarray}
Here, the first inequality is due to the convexity of the quadratic function, the second inequality is due to the concavity of the square root function, and we have used the normalization $\sum_{x_A}{\rm Tr}\{\Pi_{\lambda_j(\varrho_A)}\Pi_{x_A}\}=\braket{\lambda_j(\varrho_A)|\lambda_j(\varrho_A)}=1$ to get the last line. Notice that the last line is independent of the choice of the rank-1 PVM $\{\Pi_{x_A}\}$ or the associated orthonormal basis $\{\ket{x_A}\}$, and only depends on the Schmidt coefficients $\{\lambda_j(\varrho_A)^{1/2}\}$ of the pure bipartite state $\ket{\psi}_{AB}$. Noting this, inserting Eq. (\ref{infimum of the KD-nonreality entanglement 1}) into Eq. (\ref{proof of Proposition 1 step 4}), we thus have  
\begin{eqnarray}
&&E_{\rm KD}^{\rm NRe}(\ket{\psi}\bra{\psi}_{AB})\nonumber\\
&=&\inf_{\{\Pi_{x_A}\}\in\mathcal{M}_{\rm r1PVM}(\mathcal{H}_A)}\sum_{x_A}\big({\rm Pr}(x_A|\varrho_A)-{\rm Pr}(x_A|\varrho_A)^2\big)^{1/2}\nonumber\\
&\ge&\sum_j\big(\lambda_j(\varrho_A)-\lambda_j(\varrho_A)^2\big)^{1/2}. 
\label{proof of Proposition 1 step 5} 
\end{eqnarray}
One can then see in Eq. (\ref{proof of Proposition 1 step 5}) that the lower bound, i.e., the equality, can always be achieved by choosing $\{\Pi_{x_A}\}=\{\Pi_{\lambda_j(\varrho_A)}\}$ so that ${\rm Pr}(x_A|\varrho_A)=\lambda_{x_A}(\varrho_A)$, to obtain  
\begin{eqnarray}
E_{\rm KD}^{\rm NRe}(\ket{\psi}\bra{\psi}_{AB})&=&\sum_j\big(\lambda_j(\varrho_A)-\lambda_j(\varrho_A)^2\big)^{1/2}\nonumber\\
&=&{\rm Tr}_A\{(\varrho_A-\varrho_A^2)^{1/2}\}. 
\label{KD entanglement as a witness for linear entropy of entanglement}
\end{eqnarray}

One can also check that for two-qubit pure state, writing the Schmidt coefficients as $\{\sqrt{\lambda_1},\sqrt{\lambda_2}\}$, where $\lambda_1+\lambda_2=1$, we have $E_{\rm KD}^{\rm NRe}(\ket{\psi}\bra{\psi}_{AB})=2\sqrt{\lambda_1\lambda_2}$ which is just the entanglement concurrence for two-qubit pure state.  
\qed

\section{Proofs of Proposition 3\label{Proofs of Proposition 2}}

Let $Y_{AB}=\sum_{y_{AB}}y_{AB}\Pi_{y_{AB}}$ be a bounded Hermitian operator on a finite-dimensional Hilbert space $\mathcal{H}_{AB}$ with the complete set of eigenprojectors $\{\Pi_{y_{AB}}\}$ and the associated spectrum of eigenvalues $\{y_{AB}\}$. Similarly, let $X_A=\sum_{x_A}x_A\Pi_{x_A}$ be a bounded Hermitian operator on the Hilbert space $\mathcal{H}_A$ with the complete set of eigenprojectors $\{\Pi_{x_A}\}$ and the associated spectrum of eigenvalues $\{x_A\}$. We first have the inequality:
\begin{eqnarray}
&&\sum_{x_A,y_{AB}}|{\rm Im}({\rm Pr}_{\rm KD}(x_A;y_{AB}|\ket{\psi}\bra{\psi}_{AB})|\nonumber\\  
&=&\frac{1}{2}\sum_{x_A,y_{AB}}|{\rm Tr}\{\Pi_{y_{AB}}[\Pi_{x_A}\otimes\mathbb{I}_B,\ket{\psi}\bra{\psi}_{AB}]\}|\nonumber\\
&=&\frac{1}{2\|X_A\|_{\infty}\|Y_{AB}\|_{\infty}}\sum_{x_A,y_{AB}}\|X_A\|_{\infty}\|Y_{AB}\|_{\infty}|{\rm Tr}\{\Pi_{y_{AB}}[\Pi_{x_A}\otimes\mathbb{I}_B,\ket{\psi}\bra{\psi}_{AB}]\}|\nonumber\\
&\ge&\frac{1}{2\|X_A\|_{\infty}\|Y_{AB}\|_{\infty}}|{\rm Tr}\{Y_{AB}[X_A\otimes\mathbb{I}_B,\ket{\psi}\bra{\psi}_{AB}]\}|, 
\label{proof of lower bound for KD-nonreality entanglement for pure state step 1} 
\end{eqnarray}
where the inequality is due to the fact that the operator norm of $O$, i.e., $\|O\|_{\infty}$, is the largest singular value of $O$. Taking the supremum over all orthonormal bases $\{\ket{y_{AB}}\}$ of the Hilbert space $\mathcal{H}_{AB}$ on both sides of Eq. (\ref{proof of lower bound for KD-nonreality entanglement for pure state step 1}), we have 
\begin{eqnarray}
&&\sum_{x_A}\sup_{\{\ket{y_{AB}}\}\in\mathcal{B}_{\rm o}(\mathcal{H}_{AB})}\sum_{y_{AB}}|{\rm Im}({\rm Pr}_{\rm KD}(x_A;y_{AB}|\ket{\psi}\bra{\psi}_{AB})|\nonumber\\  
&\ge&\frac{1}{2}\sup_{Y_{AB}\in\mathbb{H}(\mathcal{H}_{AB}|\{y_{AB}\})}\frac{|{\rm Tr}\{Y_{AB}[X_A\otimes\mathbb{I}_B,\ket{\psi}\bra{\psi}_{AB}]\}|}{\|Y_{AB}\|_{\infty}\|X_A\|_{\infty}},
\label{proof of lower bound for KD-nonreality entanglement for pure state step 2} 
\end{eqnarray}
where $\mathbb{H}(\mathcal{H}_{AB}|\{y_{AB}\})$ is the set of all Hermitian operators on $\mathcal{H}_{AB}$ having a spectrum of eigenvalues $\{y_{AB}\}$. Next, note that the left-hand side of Eq. (\ref{proof of lower bound for KD-nonreality entanglement for pure state step 2}) does not depend on the spectrum of eigenvalues of $X_A$ and $Y_{AB}$. Hence, the inequality can be strengthened as 
\begin{eqnarray}
&&\sum_{x_A}\sup_{\{\ket{y_{AB}}\}\in\mathcal{B}_{\rm o}(\mathcal{H}_{AB})}\sum_{y_{AB}}|{\rm Im}({\rm Pr}_{\rm KD}(x_A;y_{AB}|\ket{\psi}\bra{\psi}_{AB})|\nonumber\\  
&\ge&\frac{1}{2}\sup_{X_A\in\mathbb{H}(\mathcal{H}_A|\{\Pi_{x_A}\})}\Big\{\sup_{Y_{AB}\in\mathbb{H}(\mathcal{H}_{AB})}\Big\{\frac{\big|{\rm Tr}\{Y_{AB}[(X_A\otimes\mathbb{I}_B),\ket{\psi}\bra{\psi}_{AB}]\}\big|}{\|Y_{AB}\|_{\infty}\|X_A\|_{\infty}}\Big\}\Big\},
\label{proof of lower bound for KD-nonreality entanglement for pure state step 3} 
\end{eqnarray}
where $\mathbb{H}(\mathcal{H}_{AB})$ is the convex set of all Hermitian operators on $\mathcal{H}_{AB}$, and $\mathbb{H}(\mathcal{H}_A|\{\Pi_{x_A}\})$ is the convex set of all Hermitian operators on $\mathcal{H}_A$ whose complete set of eigenprojectors is $\{\Pi_{x_A}\}$. Taking the infimum over the set $\mathcal{B}_{\rm o}(\mathcal{H}_A)$ of all the orthonormal bases of $\mathcal{H}_A$ on both sides of Eq. (\ref{proof of lower bound for KD-nonreality entanglement for pure state step 3}), and noting the definition of the KD-nonreality bipartite entanglement for pure state of Eq. (\ref{KD-nonreality entanglement for pure bipartite state}), we obtain 
\begin{eqnarray}
&&E_{\rm KD}^{\rm NRe}(\ket{\psi}\bra{\psi}_{AB})\nonumber\\
&\ge&\frac{1}{2}\inf_{\{\Pi_{x_A}\}\in\mathcal{M}_{\rm r1PVM}(\mathcal{H}_A)}\sup_{X_A\in\mathbb{H}(\mathcal{H}_A|\{\Pi_{x_A}\})}\Big\{\nonumber\\
&& \sup_{Y_{AB}\in\mathbb{H}(\mathcal{H}_{AB})}\Big\{\frac{\big|{\rm Tr}\{Y_{AB}[(X_A\otimes\mathbb{I}_B),\ket{\psi}\bra{\psi}_{AB}]\}\big|}{\|Y_{AB}\|_{\infty}\|X_A\|_{\infty}}\Big\}\Big\}. 
\label{proof of lower bound for KD-nonreality entanglement for pure state 4}
\end{eqnarray} 

Next, note that the left-hand side of Eq. (\ref{proof of lower bound for KD-nonreality entanglement for pure state 4}) does not depend on the operator $\tilde{Y}_{AB}:=Y_{AB}/\|Y_{AB}\|_{\infty}$. Since we have $\|\tilde{Y}_{AB}\|_{\infty}=1$, the inequality in Eq. (\ref{proof of lower bound for KD-nonreality entanglement for pure state 4}) can be further strengthened as 
\begin{eqnarray}
&&E_{\rm KD}^{\rm NRe}(\ket{\psi}\bra{\psi}_{AB})\nonumber\\
&\ge&\frac{1}{2}\inf_{\{\Pi_{x_A}\}\in\mathcal{M}_{\rm r1PVM}(\mathcal{H}_A)}\sup_{X_A\in\mathbb{H}(\mathcal{H}_A|\{\Pi_{x_A}\})}\Big\{\nonumber\\
&&\sup_{\tilde{Y}_{AB}\in\mathbb{O}(\mathcal{H}_{AB}|\|\cdot\|_{\infty}\le 1)}\Big\{\frac{\big|{\rm Tr}\{\tilde{Y}_{AB}[(X_A\otimes\mathbb{I}_B),\ket{\psi}\bra{\psi}_{AB}]\}\big|}{\|X_A\|_{\infty}}\Big\}\Big\}, 
\label{proof of lower bound for KD-nonreality entanglement for pure state 5}
\end{eqnarray}
where $\mathbb{O}(\mathcal{H}_{AB}|\hspace{1mm}\|\cdot\|_{\infty}\le 1)$ is the set of all bounded operators on $\mathcal{H}_{AB}$ with an operator norm that is less than or equal to 1. We note that the term on the right-hand side of Eq. (\ref{proof of lower bound for KD-nonreality entanglement for pure state 5}) inside the inner curly brackets is just the variational expression of the Schatten $p=1$ norm (trace norm) in terms of its conjugate $p_*=\infty$ norm (operator norm), i.e., $\sup_{\tilde{Y}_{AB}\in\mathbb{O}(\mathcal{H}_{AB}|\|\cdot\|_{\infty}\le 1)}\Big\{|{\rm Tr}\{\tilde{Y}_{AB}[(X_A\otimes\mathbb{I}_B),\ket{\psi}\bra{\psi}_{AB}]\}|\Big\}=\|[(X_A\otimes\mathbb{I}_B),\ket{\psi}\bra{\psi}_{AB}]\|_1$ \cite{Watrous book quantum information theory}. Inserting this into Eq. (\ref{proof of lower bound for KD-nonreality entanglement for pure state 5}), we obtain 
\begin{eqnarray}
E_{\rm KD}^{\rm NRe}(\ket{\psi}\bra{\psi}_{AB})&\ge&\inf_{\{\Pi_{x_A}\}\in\mathcal{M}_{\rm r1PVM}(\mathcal{H}_A)}\sup_{X_A\in\mathbb{H}(\mathcal{H}_A|\{\Pi_{x_A}\})}\|[X_A\otimes\mathbb{I}_B,\ket{\psi}\bra{\psi}_{AB}]\|_1/2\|X_A\|_{\infty}\nonumber\\
&=&\inf_{\{\Pi_{x_A}\}\in\mathcal{M}_{\rm r1PVM}(\mathcal{H}_A)}\sup_{X_A\in\mathbb{H}(\mathcal{H}_A|\{\Pi_{x_A}\})}\Delta_{X_A\otimes\mathbb{I}_B}(\ket{\psi}\bra{\psi}_{AB})/\|X_A\|_{\infty}\nonumber\\
&=&\inf_{\{\Pi_{x_A}\}\in\mathcal{M}_{\rm r1PVM}(\mathcal{H}_A)}\sup_{X_A\in\mathbb{H}(\mathcal{H}_A|\{\Pi_{x_A}\})}\Delta_{X_A}(\varrho_A)/\|X_A\|_{\infty},
\label{KD nonreality is lower bounded by a normalized trace-norm asymmetry appendix}
\end{eqnarray}
where, in the second line, we have made use of the known equality between the trace-norm noncommutativity of a Hermitian operator and pure state with the quantum standard deviation of Eq. (\ref{trace-norm asymmetry is equal to the quantum standard deviation}), and in the third line we have used $\varrho_A={\rm Tr}_B\{\ket{\psi}\bra{\psi}_{AB}\}$. 

To prove that the lower bound depends only on the Schmidt coefficients of $\ket{\psi}_{AB}$ or the eigenvalues of $\varrho_A$, it is sufficient to show that the right-hand side of Eq. (\ref{KD nonreality is lower bounded by a normalized trace-norm asymmetry appendix}) is invariant when the state is transformed under any local unitary: $\ket{\psi}\bra{\psi}_{AB}\mapsto (U_A\otimes U_B)\ket{\psi}\bra{\psi}_{AB}(U_A\otimes U_B)^{\dagger}$. Indeed, making such a state transformation on the right-hand side of Eq. (\ref{KD nonreality is lower bounded by a normalized trace-norm asymmetry appendix}) we have 
\begin{eqnarray}
&&\inf_{\{\Pi_{x_A}\}\in\mathcal{M}_{\rm r1PVM}(\mathcal{H}_A)}\sup_{X_A\in\mathbb{H}(\mathcal{H}_A|\{\Pi_{x_A}\})}\Delta_{X_A}(U_A\varrho_AU_A^{\dagger})/\|X_A\|_{\infty}\nonumber\\
\label{lower bound for KD-nonreality entanglement is local unitarily covariant 1}
&=&\inf_{\{\Pi_{x_A}\}\in\mathcal{M}_{\rm r1PVM}(\mathcal{H}_A)}\sup_{X_A\in\mathbb{H}(\mathcal{H}_A|\{\Pi_{x_A}\})}\Delta_{U_A^{\dagger}X_AU_A}(\varrho_A)/\|U_A^{\dagger}X_AU_A\|_{\infty}\\
\label{lower bound for KD-nonreality entanglement is local unitarily covariant 2}
&=&\inf_{\{\Pi_{x_A}\}\in\mathcal{M}_{\rm r1PVM}(\mathcal{H}_A)}\sup_{X'_A\in\mathbb{H}(\mathcal{H}_A|\{\Pi_{x_A}\})}\Delta_{X'_A}(\varrho_A)/\|X'_A\|_{\infty}\\
&=&\inf_{\{\Pi_{x_A}\}\in\mathcal{M}_{\rm r1PVM}(\mathcal{H}_A)}\sup_{X_A\in\mathbb{H}(\mathcal{H}_A|\{\Pi_{x_A}\})}\Delta_{X_A}(\varrho_A)/\|X_A\|_{\infty}.
\label{lower bound for KD-nonreality entanglement is local unitarily covariant 3}
\end{eqnarray}
Here, to get Eq. (\ref{lower bound for KD-nonreality entanglement is local unitarily covariant 1}), we have made use of the fact that $\|U_A^{\dagger}X_AU_A\|_{\infty}=\|X_A\|_{\infty}$, and to get Eq. (\ref{lower bound for KD-nonreality entanglement is local unitarily covariant 2}) we have defined a new Hermitian operator as $X'_A=U_A^{\dagger}X_AU_A$ and noted that $\inf_{\{\Pi_{x_A}\}\in\mathcal{M}_{\rm r1PVM}(\mathcal{H}_A)}\sup_{X'_A\in\mathbb{H}(\mathcal{H}_A|\{\Pi_{x_A}\})}\{\cdot\}=\inf_{\{\Pi_{x_A}\}\in\mathcal{M}_{\rm r1PVM}(\mathcal{H}_A)}\sup_{X_A\in\mathbb{H}(\mathcal{H}_A|\{\Pi_{x_A}\})}\{\cdot\}$ so that we get Eq. (\ref{lower bound for KD-nonreality entanglement is local unitarily covariant 3}).  
\qed

\section{Proof of Proposition 4\label{proof of proposition 4}}

Following the general result reported in Ref. \cite{Johansen quantum state from successive projective measurement}, the imaginary part of the KD quasiprobability ${\rm Pr}_{\rm KD}(x_A;y_{AB}|\ket{\psi}\bra{\psi}_{AB})$ can be expressed in terms of the post-measurement state of the nonselective binary measurement $\{\Pi_{x_A}\otimes\mathbb{I}_B,\mathbb{I}-(\Pi_{x_A}\otimes\mathbb{I}_B)\}$ over $\ket{\psi}_{AB}$, i.e., $\varrho_{AB;x_A}:=(\Pi_{x_A}\otimes\mathbb{I}_B)\ket{\psi}\bra{\psi}_{AB}(\Pi_{x_A}\otimes\mathbb{I}_B)+(\mathbb{I}-(\Pi_{x_A}\otimes\mathbb{I}_B))\ket{\psi}\bra{\psi}_{AB}(\mathbb{I}-(\Pi_{x_A}\otimes\mathbb{I}_B))$, as 
\begin{eqnarray}
&&{\rm Im}\{{\rm Pr}_{\rm KD}(x_A;y_{AB}|\ket{\psi}\bra{\psi}_{AB})\}\nonumber\\
&=&\frac{1}{2}{\rm Tr}\{(\ket{\psi}\bra{\psi}_{AB}-\varrho_{AB;x_A})e^{i(\Pi_{x_A}\otimes\mathbb{I}_B)\pi/2}\Pi_{y_{AB}}e^{-i(\Pi_{x_A}\otimes\mathbb{I}_B)\pi/2}\}. 
\end{eqnarray}
Hence, noting that $\{\Pi_{y_{AB}}\}$ is a rank-1 PVM basis, using Eq. (\ref{Lemma on the variational expression on the trace-norm}) of Lemma 1 we obtain 
\begin{eqnarray}
&&E_{\rm KD}^{\rm NRe}(\ket{\psi}\bra{\psi}_{AB})\nonumber\\
&=&\inf_{\{\ket{x_A}\}\in\mathcal{B}_{\rm o}(\mathcal{H}_A)}\sum_{x_A}\sup_{\{\ket{y_{AB}}\}\in\mathcal{B}_{\rm o}(\mathcal{H}_{AB})}\sum_{y_{AB}}|{\rm Im}({\rm Pr}_{\rm KD}(x_A;y_{AB}|\ket{\psi}\bra{\psi}_{AB})|\nonumber\\
&=&\inf_{\{\Pi_{x_A}\}\in\mathcal{M}_{\rm r1PVM}(\mathcal{H}_A)}\sum_{x_A}\nonumber\\
&\cdot&\sup_{\{\Pi_{y_{AB}}\}\in\mathcal{M}_{\rm r1PVM}(\mathcal{H}_{AB})}\sum_{y_{AB}}\frac{1}{2}{\rm Tr}\{e^{-i(\Pi_{x_A}\otimes\mathbb{I}_B)\pi/2}(\ket{\psi}\bra{\psi}_{AB}-\varrho_{AB;x_A})e^{i(\Pi_{x_A}\otimes\mathbb{I}_B)\pi/2}\Pi_{y_{AB}}\}\nonumber\\
&=&\inf_{\{\Pi_{x_A}\}\in\mathcal{M}_{\rm r1PVM}(\mathcal{H}_A)}\sum_{x_A}\frac{1}{2}\|e^{-i(\Pi_{x_A}\otimes\mathbb{I}_B)\pi/2}(\ket{\psi}\bra{\psi}_{AB}-\varrho_{AB;x_A})e^{i(\Pi_{x_A}\otimes\mathbb{I}_B)\pi/2}\|_1\nonumber\\
&=&\inf_{\{\Pi_{x_A}\}\in\mathcal{M}_{\rm r1PVM}(\mathcal{H}_A)}\sum_{x_A}\frac{1}{2}\|\ket{\psi}\bra{\psi}_{AB}-\varrho_{AB;x_A}\|_1, 
\end{eqnarray}
where in the last line we have made use of the fact that the trace distance is invariant under unitary transformation. Hence, we have proved Eq. (\ref{KD-nonreality entanglement as trace-norm of local measurement disturbance}) of Proposition 4. It is easy to check that when the pure bipartite state is factorizable, $\ket{\psi}_{AB}=\ket{\psi_A}_A\ket{\psi_B}_B$, then by choosing a rank-1 PVM $\{\Pi_{x_A}\}$, one of whose elements is $\ket{\psi_A}\bra{\psi_A}$, we have $\varrho_{AB;x_A}=\ket{\psi}\bra{\psi}_{AB}$ for all $x_A$, so that $E_{\rm KD}^{\rm NRe}(\ket{\psi}\bra{\psi}_{AB})=0$, as expected. 

Finally, the inequality in Eq. (\ref{KD-nonreality entanglement is upper bounded by trace-norm of local nonselective measurement disturbance}) is due to the fact that trace distance is nonincreasing under a partial trace, and that ${\rm Tr}_B\{\varrho_{AB;x_A}\}=\Pi_{x_A}\varrho_A\Pi_{x_A}+(\mathbb{I}_A-\Pi_{x_A})\varrho_A(\mathbb{I}_A-\Pi_{x_A})=\varrho_{A;x_A}$. 
\qed

\section{Proof of Proposition 5 \label{Proof of Theorem 2}}

First, the upper bound in Eq. (\ref{upper and lower bounds for the KD-nonreality entanglement for generic state}) can be obtained as
\begin{eqnarray}
E_{\rm KD}^{\rm NRe}(\varrho_{AB})&=&\inf_{\substack{\{p_k,\ket{\psi^k}\bra{\psi^k}_{AB}\}\\ \varrho_{AB}=\sum_kp_k\ket{\psi^k}\bra{\psi^k}_{AB}}}\sum_k p_kE_{\rm KD}^{\rm NRe}(\ket{\psi^k}\bra{\psi^k}_{AB})\nonumber\\
\label{Proof of upper bound in Theorem 2 step 2}
&=&\inf_{\substack{\{p_k,\ket{\psi^k}\bra{\psi^k}_{AB}\}\\ \varrho_{AB}=\sum_kp_k\ket{\psi^k}\bra{\psi^k}_{AB}}}\sum_k p_k S_{\rm KD}^{\rm NRe}(\varrho_A^k)\\
\label{Proof of upper bound in Theorem 2 step 3}
&\le&\inf_{\substack{\{p_k,\ket{\psi^k}\bra{\psi^k}_{AB}\}\\ \varrho_{AB}=\sum_kp_k\ket{\psi^k}\bra{\psi^k}_{AB}}}S_{\rm KD}^{\rm NRe}\big(\sum_k p_k \varrho_A^k\big)\\
\label{Proof of upper bound in Theorem 2 step 4}
&=&S_{\rm KD}^{\rm NRe}\big(\varrho_A\big). 
\end{eqnarray}
Here, to get Eq. (\ref{Proof of upper bound in Theorem 2 step 2}) we have used Eq. (\ref{KD-nonreality entanglement as quantum entropy of the reduced density operator}) of Proposition 2 where $\varrho_A^k={\rm Tr}_B\{\ket{\psi^k}\bra{\psi^k}_{AB}\}$, Eq. (\ref{Proof of upper bound in Theorem 2 step 3}) is obtained by virtue of the concavity of $S_{\rm KD}^{\rm NRe}\big(\varrho_A\big)$ with respect to $\varrho_A$, and Eq. (\ref{Proof of upper bound in Theorem 2 step 4}) is due to the fact that $\varrho_A={\rm Tr}_B\{\varrho_{AB}\}=\sum_kp_k{\rm Tr}_B\{\ket{\psi^k}\bra{\psi^k}_{AB}\}=\sum_kp_k\varrho_A^k$, which is valid for all the pure state decompositions of $\varrho_{AB}$. 

We proceed to prove the lower bound in Eq. (\ref{upper and lower bounds for the KD-nonreality entanglement for generic state}). First, let us assume that the state admits a pure state decomposition $\varrho_{AB}=\sum_kp_k\ket{\psi^k}\bra{\psi^k}_{AB}$. Then, from Eq. (\ref{proof of lower bound for KD-nonreality entanglement for pure state step 1}), we have 
\begin{eqnarray}
&&\sum_kp_k\sum_{x_A,y_{AB}}|{\rm Im}({\rm Pr}_{\rm KD}(x_A;y_{AB}|\ket{\psi^k}\bra{\psi^k}_{AB})|\nonumber\\  
&\ge&\frac{1}{2\|X_A\|_{\infty}\|Y_{AB}\|_{\infty}}\sum_kp_k|{\rm Tr}\{Y_{AB}[X_A\otimes\mathbb{I}_B,\ket{\psi^k}\bra{\psi^k}_{AB}]\}|\nonumber\\
&\ge&\frac{1}{2\|X_A\|_{\infty}\|Y_{AB}\|_{\infty}}|{\rm Tr}\{Y_{AB}[X_A\otimes\mathbb{I}_B,\sum_kp_k\ket{\psi^k}\bra{\psi^k}_{AB}]\}|\nonumber\\
&=&\frac{1}{2\|X_A\|_{\infty}\|Y_{AB}\|_{\infty}}|{\rm Tr}\{Y_{AB}[X_A\otimes\mathbb{I}_B,\varrho_{AB}]\}|, 
\end{eqnarray}
where we have made use of the triangle inequality and the pure states decomposition in the last two lines. Following the same arguments as in the proof of Proposition 3, we thus obtain 
\begin{eqnarray}
&&\sum_kp_k\inf_{\{\ket{x_A}\}\in\mathcal{B}_{\rm o}(\mathcal{H}_A)}\sum_{x_A}\sup_{\{\ket{y_{AB}}\in\mathcal{B}_{\rm o}(\mathcal{H}_{AB})\}}\sum_{x_A,y_{AB}}|{\rm Im}({\rm Pr}_{\rm KD}(x_A;y_{AB}|\ket{\psi^k}\bra{\psi^k}_{AB}))|\nonumber\\  
&\ge&\inf_{\{\Pi_{x_A}\}\in\mathcal{M}_{\rm r1PVM}(\mathcal{H}_A)}\sup_{X_A\in\mathbb{H}(\mathcal{H}_A|\{\Pi_{x_A}\})}\|[X_A\otimes\mathbb{I}_B,\varrho_{AB}]\|_1/2\|X_A\|_{\infty}. 
\end{eqnarray}
Taking the infimum over all possible pure states decompositions of $\varrho_{AB}$, and noting that the right-hand side is independent of such pure states decompositions and the definition of KD-nonreality entanglement for state $\varrho_{AB}$ in Eq. (12), we thus obtain
\begin{eqnarray}
E_{\rm KD}^{\rm NRe}(\varrho_{AB})&\ge&\inf_{\{\Pi_{x_A}\}\in\mathcal{M}_{\rm r1PVM}(\mathcal{H}_A)}\sup_{X_A\in\mathbb{H}(\mathcal{H}_A|\{\Pi_{x_A}\})}\|[X_A\otimes\mathbb{I}_B,\varrho_{AB}]\|_1/2\|X_A\|_{\infty}.
\end{eqnarray}
\qed

\end{document}